\documentclass[12pt]{article}
\usepackage{amsmath}
\usepackage{graphicx}
\usepackage{enumerate}
\usepackage{natbib}
\usepackage{url} % not crucial - just used below for the URL 

\allowdisplaybreaks
\newcommand{\blind}{0}

\usepackage{rotating}
\usepackage{color}

\usepackage{relsize}

\usepackage{multirow}
\def\E{\mathbb{E}}

\def\T{{\mathrm{\scriptscriptstyle T} }}

\makeatletter
\newcommand*{\indep}{%
	\mathbin{%
		\mathpalette{\@indep}{}%
	}%
}

\newcommand*{\nindep}{%
	\mathbin{% % The final symbol is a binary math operator
		\mathpalette{\@indep}{\not}% \mathpalette helps for the adaptation
	}%
}
\newcommand*{\@indep}[2]{%
	\sbox0{$#1\perp\m@th$}% box 0 contains \perp symbol
	\sbox2{$#1=$}% box 2 for the height of =
	\sbox4{$#1\vcenter{}$}% box 4 for the height of the math axis
	\rlap{\copy0}% first \perp
	\dimen@=\dimexpr\ht2-\ht4-.2pt\relax
	\kern\dimen@
	{#2}%
	\kern\dimen@
	\copy0 % second \perp
}
\makeatother
 \usepackage{booktabs}
 \usepackage{amssymb}

\newtheorem{theorem}{Theorem}

\newtheorem{proof}{Proof}

\newtheorem{proposition}{Proposition}
\newtheorem{assumption}{Assumption}

\usepackage{caption}
\usepackage{subcaption}
\usepackage{amsfonts}
\def\E{\mathbb{E}}

\usepackage{subcaption}  % 添加subcaption包
\usepackage[all]{xy}
\usepackage{bm}
\usepackage{pifont}
\usepackage{stmaryrd}
\usepackage{pgf,tikz}
\usepackage{tikz}
\usetikzlibrary{shapes,arrows,positioning,calc}
\usepackage{hyperref}
\hypersetup{
    colorlinks=true,
    linkcolor=blue,
    urlcolor=blue,
    citecolor=blue
}

\usepackage[flushleft]{threeparttable}
\usepackage{enumitem}
\usepackage{mathrsfs}

\def\pr{\textnormal{pr}}
\def\AT{{ss}}
\def\CO{{s\bar{s}}}
\def\NT{{\bar{s}\bar{s}}}
\def\DE{{\bar{s}s}}  
\usepackage{arydshln}

\def\pr{\textnormal{pr}}

\def\T{{ \mathrm{\scriptscriptstyle T} }}

\def\T{{ \mathrm{\scriptscriptstyle T} }}

\def\E{\mathcal{E}}

\usepackage{wrapfig}

\def\T{{ \mathrm{\scriptscriptstyle T} }}

\def\E{\mathbb{E}}
\def\T{{ \mathrm{\scriptscriptstyle T} }}
\def\pr{\textnormal{pr}}
\def\AT{{ss}}
\def\CO{{s\bar{s}}}
\def\NT{{\bar{s}\bar{s}}}
\def\DE{{\bar{s}s}}

\def\AT{{\mathrm{AT}}}
\def\LL{{ss}}
\def\CO{{\mathrm{CO}}}
\def\LD{{s\bar{s}}}
\def\NT{{\mathrm{NT}}}
\def\DL{{\bar{s}s}}
\def\DE{{\mathrm{DE}}}
\def\DD{{\bar{s}\bar{s}}}

\usepackage{hyperref}

\usepackage[symbol]{footmisc}

\usepackage{xr}
\makeatletter
\newcommand*{\addFileDependency}[1]{
  \typeout{(#1)}
  \@addtofilelist{#1}
  \IfFileExists{#1}{}{\typeout{No file #1.}}
}
\makeatother
 
\newcommand*{\myexternaldocument}[1]{
    \externaldocument{#1}
    \addFileDependency{#1.tex}
    \addFileDependency{#1.aux}
}
 
\myexternaldocument{SM}

\begin{document}

	\def\spacingset#1{\renewcommand{\baselinestretch}%
		{#1}\small\normalsize} \spacingset{1}

	\newcounter{savecntr}% Save footnote counter
\newcounter{restorecntr}
	%%%%%%%%%%%%%%%%%%%%%%%%%%%%%%%%%%%%%%%%%%%%%%%%%%%%%%%%%%%%%%%%%%%%%%%%%%%%%%
		\if0\blind
	{  

    	\title{\bf Causal Inference  with Outcomes Truncated by Death and Missing Not at Random}
        
 			\author{Wei Li\setcounter{savecntr}{\value{footnote}}\thanks{Both authors contributed equally to this work.}
 			\textsuperscript{1},     Yuan Liu\setcounter{restorecntr}{\value{footnote}}%
 			\setcounter{footnote}{\value{savecntr}}\footnotemark% Print footnotemark
 			\setcounter{footnote}{\value{restorecntr}} \textsuperscript{1,3},  
 			Shanshan Luo\textsuperscript{2}, and Zhi Geng\textsuperscript{2}
 %\thanks{Corresponding author}
    \\\\
 			\textsuperscript{1} Center for Applied Statistics and School of Statistics, \\Renmin University of China\\\\
 			\textsuperscript{2} School of Mathematics and Statistics, \\Beijing Technology and Business University\\\\ 			\textsuperscript{2} Mathematical Institute, \\Leiden University\\
 		}
		\date{}
		
		\maketitle
	} \fi
	
	\if1\blind
	{
		\bigskip
		\bigskip
		\bigskip
		\begin{center}
			{\LARGE\bf Efficiency-improved doubly robust estimation with non-confounding predictive covariates}
		\end{center}
		\medskip
	} \fi
		\medskip

\begin{abstract}  
In clinical trials,  principal stratification analysis is commonly employed to address the issue of truncation by death, where a subject dies before the outcome can be measured. However, in practice, many survivor outcomes may remain uncollected or be missing not at random, posing a challenge to standard principal stratification analysis.  In this paper, we explore the identification, estimation, and bounds of the average treatment effect within a subpopulation of individuals who would potentially survive under both treatment and control conditions. We show that the causal parameter of interest can be identified by introducing a proxy variable that affects the outcome only through the principal strata, while requiring that the treatment variable does not directly affect the missingness mechanism. Subsequently, we propose an approach for estimating causal parameters and derive nonparametric bounds in cases where identification assumptions are violated. We illustrate the performance of the proposed method through simulation studies and a real dataset obtained from a Human Immunodeficiency Virus (HIV) study.
\end{abstract} 
 \bigskip
 \bigskip
 \bigskip
\noindent%
\vfill

\newpage
\spacingset{2} % DON'T change the spacing!
  \section{Introduction}   
Researchers are typically interested in assessing risk factors for outcome variables in clinical trials. However, in long-term trials, some subjects may die before the follow-up assessment, resulting in undefined outcomes, a problem known as truncation by death \citep{ding2011identifiability,wang2017identification,wang2017causal}. Additionally, even if some subjects survive, their outcomes may still not be collected or might be missing not at random (MNAR). For instance, in a study on the treatment effectiveness for patients with the human immunodeficiency virus type I  (HIV-I), researchers randomly assigned these patients to two competing antiretroviral therapies \citep{gilbert2003sensitivity}. Suppose we are interested in evaluating the relative effects of these treatments on the CD4 levels of these patients after approximately two years. In this long-term clinical trial, some patients died before their final CD4 levels could be measured, leading to truncation by death. Moreover, some survivors with poor health conditions did not participate in the follow-up survey, resulting in missing outcomes. 
% \citep{little2019statistical}.

Two potential problems arise when directly comparing the observed outcomes for survivors of two different treatment regimes. Firstly, survivors may come from different latent subpopulations, making such direct comparisons difficult to interpret causally  \citep{frangakis2002principal,rubin2006causal}. For example, there might be a subgroup of individuals who would survive under any treatment regime and another group who can only survive in the control group. This implies that the survivors in the treatment and control groups are potentially not from the same population \citep{frangakis2002principal}. Secondly, whether the outcomes (i.e., CD4 levels) of survivors are missing may depend on the outcomes themselves. For example, individuals in poorer physical condition may drop out during the follow-up study, leading to missing outcome variables. This can result in selection bias if the analysis is based solely on the complete data \citep{little2019statistical}.

When  the outcomes of survivors are completely observed, most previous studies have employed the principal stratification framework to address the truncation by death problem. \cite{robins1986new} proposed estimating the average causal effect within the always-survivor group, which comprises subjects who would survive regardless of the treatment they receive. The contrast within the always-survivor group was later termed the survivor average causal effect \citep{frangakis2002principal}, which  is causally interpretable because membership in the always-survivor group is defined at baseline.
Without further assumptions, the survivor average causal effect cannot be identified. \cite{zhang2003estimation} derived large sample bounds for causal effects within the principal strata.  
Several works such as \cite{gilbert2003sensitivity} and \cite{lee2010causal} performed sensitivity analysis to evaluate principal causal effects. However, both the bound and sensitivity analysis approaches are not precise enough to provide definitive conclusions. Some scholars have considered the identification problem of principal causal effects \citep{ding2011identifiability,wang2017identification,wang2017causal,chen2009identifiability,zhang2009likelihood,jiang2016principal,luo2023causal,deng2024direct,wang2024causal,luo2024causal}. Specifically, 
\cite{zhang2009likelihood} developed a likelihood-based approach for principal causal effects under Gaussian mixture models.
\cite{ding2011identifiability} established the nonparametric identification of principal causal effects using pretreatment covariates, where the distribution of these covariates provides information about the always-survivor group. %This approach was later extended to observational studies by \citet{wang2017identification}. 
\cite{wang2017identification}  relaxed the identification assumptions in  \cite{ding2011identifiability}  by incorporating more detailed covariate information for principal stratification.

Although numerous methods have been developed for truncation by death problem, they cannot be directly applied to our motivating example, where the outcomes of many survivors are missing and likely to be missing not at random \citep{chen2009identifiability,ding2018causal,sun2018semiparametric,sun2018inverse,miao2024identification,liu2020identification,shi2023data,li2023non}. 
%However, in the motivating example we are concerned with,  the outcomes of many survivors are missing and likely to be missing not at random. 
This is because survivors may choose not to participate in follow-up studies due to poor health, such as low CD4 levels, resulting in missing CD4 data. Few studies have simultaneously considered truncation by death and missing not at random problems when evaluating principal causal effects except a recent work by \cite{bia2022assessing}. %\citet{bia2022assessing} did not consider identifiability issues. Instead, they
They introduced structural assumptions about the observational data and employed Bayesian approaches for inference with continuous outcomes. While flexible and efficient, Bayesian methods rely on subjective prior assumptions, posing challenges for selecting appropriate priors in our motivating example with binary outcomes. 

In this paper, we consider the identification, estimation, and bounds of the survivor average treatment effect when the binary outcomes are both truncated by death and missing not at random. Under the introduced treatment-independent missingness setting, we use a proxy variable that affects the outcome only through the principal strata to establish identification. The treatment-independent missingness assumption requires that the treatment variable does not directly influence the missingness mechanism. This assumption is closely related to the commonly used shadow variable assumption in the MNAR literature \citep{li2023non,miao2016varieties}. Based on the identification results, we then provide  a procedure  to
estimate the survivor average causal effect. 
We also examine potential violations of our identification model, providing nonparametric bounds for the survivor average causal effect. Our proposed approach applies to both randomized and observational studies, and accommodates measured common causes of the survival and outcome processes.  
The finite sample properties of the proposed estimators are assessed through simulation studies, and we apply our approach to analyze a real dataset from the HIV study.

The rest of the paper is organized as follows. In Section \ref{sec:iden}, we introduce notation, definitions, and assumptions, while also presenting identification results for the causal parameter of interest. In Section \ref{Sec:Methodology}, we develop an estimation approach for the survivor average treatment effect.  In Section \ref{sec:bounds}, we provide the nonparameteric bound results when some identification assumptions are violated. Section \ref{sec:sim} presents simulation studies to evaluate the finite-sample performance of the proposed approach, followed by an application to an HIV dataset in Section \ref{sec:application}. We end with a discussion in Section \ref{sec:dis}. Proofs of theorems and propositions are relegated to the supplementary material.

\section{Notation, Assumptions, and Identification}
\label{sec:iden}
\subsection{Setup}

In our study, we have a sample of $ n $ patients with the human immunodeficiency virus type I (HIV-I), indexed by $ i = 1, \ldots, n $. These $ n $ patients are considered a random sample drawn from a large super-population. For each patient $i$, let $ X_i $ denote a vector of baseline covariates. Let $ Z_i $ represent a binary treatment variable, where $ Z _i = 0 $ indicates assignment to zidovudine-only therapy (control group, abbreviated as ZDV-only regime), while $ Z_i = 1 $ if assigned to the new therapies including zidovudine plus didanosine, zidovudine plus zalcitabine and didanosine alone (treatment group). 
Researchers also collected the following measurements after 96 weeks of follow-up: a binary variable $ S_i $ representing the survival status, where $ S_i = 0 $ if individuals die before 96 weeks, and $ S_i = 1 $ if they survive; and a binary outcome variable $ Y_i $ indicating whether the CD4 T cell counts at 96 weeks are higher than at baseline, where $ Y _i= 1 $ indicates higher counts and $ Y _i= 0 $ otherwise. Additionally, some survivors may not participate in the follow-up survey. Therefore, we introduce a missingness indicator $ R_i $ at 96 weeks for survivors. Here, $ R_i = 1 $ indicates that a survivor's CD4 level is observed, while $ R _i= 0 $ indicates that the outcome is missing. For patients who died, i.e., $S_i=0$, we always set the missingness indicator $ R_i = 0 $. In fact, it is only possible to collect the outcome $ Y _i$ for individuals with $ S_i = R_i = 1 $.  There are only three possible values of $ (S _i, R _i)$ in the data, $ (S_i, R_i) = (0, 0)$, $ (S_i, R_i) = (1, 0)$ and $ (S_i, R_i) = (1, 1)$.  For simplicity, we omit the subscript $ i $ in the following discussion. 

\begin{table}[t]
	\centering 
	\caption{Data structure, where ‘$\checkmark$’ denotes the observable data,  ‘$\times$’     denotes the missing data, and `NA' denotes the undefined data.}
	\resizebox{0.5949595\textwidth}{!}{   
		\begin{tabular}{cccccccccccccc}
			\toprule\addlinespace[0.5mm]
			\multicolumn{6}{c}{Without monotonicity assumption \ref{assu:mono}} &  &  & \multicolumn{6}{c}{With monotonicity assumption \ref{assu:mono}} \\ \addlinespace[1mm]\cline{1-6} \cline{9-14} \addlinespace[1mm]
			$X$           & $Z$       & $S$       & $R$       & $Y$           & $G$              &  &  & $X$           & $Z$      & $S$      & $R$      & $Y$           & $G$              \\\addlinespace[0.5mm]
			$\checkmark$       & $0$       & 0         & 0         & NA      & $ \NT,\CO $      &  &  & $\checkmark$       & $0$      & 0        & 0        & NA       & $ \NT,\CO $      \\
			$\checkmark$       & $1$       & 0         & 0         & NA     & $ \NT,\DE $      &  &  & $\checkmark$       & $1$      & 0        & 0        & NA       & $ \NT $          \\
			$\checkmark$       & $0$       & 1         & 0         & $\times$      & $ \AT,\DE $      &  &  & $\checkmark$       & $0$      & 1        & 0        & $\times$      & $ \AT $          \\
			$\checkmark$       & $1$       & 1         & 0         & $\times$      & $ \AT,\CO $      &  &  & $\checkmark$       & $1$      & 1        & 0        & $\times$      & $ \AT,\CO $      \\
			$\checkmark$       & $0$       & 1         & 1         & $\checkmark$       & $ \AT,\DE $      &  &  & $\checkmark$       & $0$      & 1        & 1        & $\checkmark$       & $ \AT $          \\
			$\checkmark$       & $1$       & 1         & 1         & $\checkmark$       & $ \AT,\CO $      &  &  & $\checkmark$       & $1$      & 1        & 1        & $\checkmark$       & $ \AT,\CO $      \\ \addlinespace[1mm]\bottomrule
	\end{tabular}} 
	\label{tab:descrip-data} 
\end{table}

We use potential outcomes to define causal effects \citep{rubin1974estimating,rubin1978bayesian}. Let $S (z) $   and $Y (z)$ denote the potential survival status and the potential change of  CD4 T cell counts under the treatment level $Z=z$, respectively. We make the stable unit treatment value assumption (SUTVA), which means there is no interference between units and there is only one version of potential outcomes for one unit   \citep{rubin1990comment}.   In the potential outcomes framework, {average causal effects are defined as the difference in the expectations of potential outcomes between the treatment and control group.}  However, 
defining the causal effects of the treatment on the change of CD4 T cell counts in our study requires accounting for the outcome being truncated by death. Specifically, because there is no clear definition of the potential outcome $ Y(z) $ when $ S(z) = 0 $, direct comparisons of potential outcomes may not be appropriate. We address this issue using the principal stratification framework. 

In our study, based on the binary treatment and the binary survival status, we classify HIV patients into four (latent) principal strata using the joint potential values $ \{S_{ }(1), S_{ }(0)\}$. The always-survivors   $\LL=\{ i:S_{i }(1)= S_{i }(0)=1\}$: subjects who would survive regardless of the treatment they receive. The never-survivors   $\DD=\{ i:S_{i }(1)= S_{i }(0)=0\}$:  subjects who would die regardless of the treatment they receive. The compliers   ${\LD}=\{ i:S_{i }(1)=1, S_{i }(0)=0\}$:  subjects who would survive if they receive the new therapies, but would die if they receive ZDV-only regime. The defiers ${\DL}=\{ i:S_{i }(1)=0, S_{i }(0)=1\}$:  subjects who would die if they receive the new therapies, but would survive if they receive the ZDV-only regime.
Let $G$ denote the baseline principal stratum membership, and $G\in\{\LL, {\DD},{\LD},{\DL}\}$. The first part of Table \ref{tab:descrip-data} shows the possible data structure.

Principal causal effects are defined as the population average causal effects conditional on the principal strata $G$.  For each individual, we cannot observe the membership of the principal stratum $G$, as we cannot observe $S(0)$ and $S(1)$ for any individual simultaneously. For subjects belonging to the principal stratum $G=\LD$, $ Y (0) $  is undefined; for subjects belonging to the principal stratum $G=\DL$, $ Y (1) $  is undefined; and for subjects belonging to the principal stratum $G=\DD$, there are no  CD4 T cell counts available for comparison. These analyses indicate that a well-defined real value for the average causal effect of ZDV-only versus the new therapies exists only for the always survivor  group  ($G=\LL$). Following   \cite{rubin2006causal}, we define the survivor average causal effect (SACE) as: 
$ 
\Delta_{\LL}=\mathbb{E}\{Y(1)-Y(0) \mid G=\LL\}.
$
\subsection{Assumptions and identification} 
\label{sec:assumption}
We first introduce some commonly used assumptions in the causal inference literature.

\begin{assumption}[Strong ignorability and overlap] 
	\label{assu:igno}
	(i) $
	Z \indep \{Y(1),Y(0),S(1),S(0)\}\mid {X}
	$; (ii) $0<\pr(Z=1\mid X)<1$.
\end{assumption} 

Assumption \ref {assu:igno} implies that within units defined by the pretreatment covariates  $X$, the treatment is randomly assigned, with the probability depending on the values of covariates.   In our application, Assumption \ref {assu:igno}(i)  naturally holds because the treatment assignment mechanism is randomized. We also require the overlap assumption  \ref {assu:igno}(ii), which requires sufficient overlap in the joint distribution of covariates between the treatment and control groups.  Assumption \ref{assu:igno} is widely used in principal stratification analysis when the outcome variable is fully observed for survivors \citep{ding2011identifiability,wang2017identification,jiang2016principal, luo2023causal}.
It has also been introduced to describe the treatment assignment mechanism in \cite{bia2022assessing} when outcomes are both truncated by death and missing not at random.   
{Let $\mu_{zg}(X) $ denote the conditional probability $\pr(Y=1\mid Z=z, G=g,X)$. Given Assumption \ref{assu:igno}, the target estimand \(\Delta_{\LL} = \mathbb{E}\{Y(1) - Y(0) \mid G = \LL\}\) can be expressed  as:
	\begin{align}\label{eqn:deltass}
		\Delta_{\LL} &= \frac{\mathbb{E}\{ \mu_{1\LL}(X)  \pr(G = \LL \mid X) \}}{\mathbb{E}\{ \pr(G = \LL \mid X) \}}   - \frac{\mathbb{E}\{ \mu_{0\LL}(X)  \pr(G = \LL \mid X) \}}{\mathbb{E}\{ \pr(G = \LL \mid X) \}}.
	\end{align}
	In the expression \eqref{eqn:deltass},
	the numerator of each term corresponds to the weighted average of the conditional probability $\mu_{z\LL}(X) $ ($z=1,0$), with weights given by the probability of belonging to stratum \(\LL\). The denominator, \(\mathbb{E}\{\pr(G = \LL \mid X)\}=\pr(G = \LL)\), serves as a normalizing constant. It ensures the resulting contrast between the two terms represents the average causal effect within the target subpopulation defined by \(G=\LL\). This weighting structure is analogous to the SATT formula (Equation 5)  in \cite{bia2022assessing}, where the denominator captures the empirical support over which the estimand is defined.
	In the following, we impose additional assumptions to guarantee that both $\mu_{z\LL}(X)$ and $\pr(G=\LL\mid X)$ in \eqref{eqn:deltass} are identifiable from the observed data distribution.}

To nonparametrically identify {the proportions of different principal strata, \(\pr(G = g \mid X)\),} we impose the following restriction on the survival behaviors. 
\begin{assumption}[Monotonicity]
	
	$
	S (1)\geq S (0).
	$
	\label{assu:mono}
\end{assumption}

Assumption \ref{assu:mono} is commonly employed in principal stratification analysis \citep{ding2011identifiability, wang2017identification}. This assumption suggests that the treatment does not have a negative effect on the survival status of any individual, thus ruling out the presence of the defier subgroup $G=\DL$. Assumption \ref{assu:mono} is reasonable in our empirical application, as trials in medical research typically offer new treatment regimes that are at least as effective as the current standard \citep{djulbegovic2012new}, and the new treatment regime has been verified to extend patients' survival time  \citep{englund1997zidovudine}. The validity of 
monotonicity cannot be directly tested, but this assumption imposes testable restrictions on the
probability distribution of observed data in certain cases. 
By separately calculating the survival rates for the treatment and control groups, we find that the observed data distributon in our study does not contradict the monotonicity
assumption. The second part of Table \ref{tab:descrip-data} shows the possible data structure under monotonicity assumption  \ref{assu:mono}. Under Assumptions \ref{assu:igno} and \ref{assu:mono}, we can identify the proportions of different principal strata in the population. Specifically, we have,
\begin{equation}
	\label{eq:mon}
	\begin{gathered}
		\pr(G = \DD\mid X)  =\pr(S = 0\mid Z=1,X),~
		\pr(G = \LL\mid X)    = \pr(S = 1\mid Z = 0,X),\\
		\pr(G = \LD\mid X)  =1 -  \pr(G = \DD\mid X)-  \pr(G = \LL\mid X).
	\end{gathered}
\end{equation}

We introduce the following treatment-independent missingness assumption to account for the issue of outcomes missing not at random  for  some survivors.
\begin{assumption}[Treatment-independent missingness]
	(i) $ Z  \nindep Y \mid (S=1, {X})$, (ii) $Z \indep R  \mid (S = 1, {X},Y).$  
	\label{assu:shadow}
\end{assumption}

Assumption \ref{assu:shadow}(i) implies that, conditional on the covariates among survivors, treatment assignment $ Z $ is associated with outcome $ Y $. As validated in medical research \citep{englund1997zidovudine, trialists1999zidovudine}, the new HIV treatment regime exhibits significant differences compared to the original ZDV-only therapy, thus Assumption \ref{assu:shadow}(i) is reasonable in our context. 
Assumption \ref{assu:shadow}(ii) requires that, given the covariates and outcomes among survivors, treatment assignment $ Z $ is conditionally independent of the missingness indicator $ R $. Studies have shown that missing data in long-term clinical trials are often associated with patients' underlying health conditions and may not be related to the treatment assignments \citep{twisk2002attrition,sterne2009multiple}. In our example,  patients in poorer health, potentially indicated by lower CD4 levels, are more likely to drop out of the study and might be independent of the treatment received.  
Assumption \ref{assu:shadow} is also linked to the shadow variable assumption in missing data analysis   and is widely employed to address missing not at random problems \citep{li2023non,miao2016varieties,yang2019causal}. 

While  Assumption \ref{assu:shadow}  can be used to identify the conditional probability $\pr(Y=1\mid Z,X,S=1)$ when $Y$ is missing not at random, which represents the distribution of outcomes among the survivors, it is not sufficient to guarantee identification of the parameter $\Delta_{\LL}$. Therefore, we further introduce a baseline covariate to  capture the information from the latent principal stratification, similar to the approach used in principal stratification analysis with complete data \citep{wang2017identification}.   
Assume that baseline covariates $X$ can be written as $(C^\T, A)^\T$. Here, $A$ is a scalar covariate that can affect the outcome only through the principal strata, conditional on the treatment assignment $Z$ and the remaining covariates $C$ among the survivors. For convenience, the notation  $X$  and $(C^\T,A)^\T$	 may be used interchangeably below. Specifically, we make the following assumption.
\begin{assumption}[Proxy variable] (i) $
	A \indep Y\mid ({Z=1},G,C)$, (ii) $A \nindep G\mid (Z=1,S=1,C). %%%确认一下
	$
	\label{assu:proxy}
\end{assumption}

Assumption \ref{assu:proxy}(i) means that $A$ has no direct effect on outcome $Y$, and $ A $ affects outcome $ Y $ through its effect on other variables  $ (Z,G,C )$. % and $A$ should be correlated with principal strata $G$. 
{Assumption \ref{assu:proxy}(ii) is similar to the relevance assumption in instrumental variable analysis. The equations in \eqref{eq:mon} imply that $\pr(G=\AT \mid Z=1, S=1, X)$ is identifiable through the expression:  
	$
	\pr(G=\AT \mid Z=1, S=1, X) = {\pr(G=\AT \mid X)}/{\{\pr(G=\AT \mid X) + \pr(G=\CO \mid X)\}}.$ 
	Since $X=(C^\T,A)^\T$, we can integrate over $A$ to obtain the identifiability of
	$\pr(G=\AT \mid Z=1, S=1, C)$. By comparing $\pr(G=\AT \mid Z=1, S=1, X)$ with $\pr(G=\AT \mid Z=1, S=1, C)$, we can empirically assess the validity of Assumption \ref{assu:proxy}(ii).} 
Similar assumptions are widely used in principal stratification analysis \citep{ding2011identifiability,wang2017identification}.  
The baseline covariate 
$A$ contains information on the latent principal stratification $G$, which makes identifying the principal causal effects possible and thus is called a proxy or substitutional variable. 
In our example, $A$ is the  CD4 level of a patient at baseline, which is an important physical condition indicator, especially in HIV patients. The principal stratification $G$ characterizes the physical condition of the patient approximately. To illustrate, patients in the always-survivor group ($G = \LL$) are, with high probability, in a physically superior state compared to those in the never-survivor group ($G = \DD$).   Therefore, it is reasonable to consider $A$ as a proxy for $G$, and  Assumption  \ref{assu:proxy}(ii) is empirically valid. Furthermore, the outcome  $Y$ is the change of CD4 level after the treatment, which is
usually believed to be not directly associated with the baseline CD4 level, but may be
associated with the treatment and the physical condition of a patient, so  Assumption  \ref{assu:proxy}(i)  is also reasonable here. 
The simplest causal graph associated with Assumptions \ref{assu:igno}-\ref{assu:proxy} is provided in Figure \ref{fig:causal}. {Besides, we have extended our theoretical analysis to incorporate intermediate variables that account for how  $A$ affects $Y$ through
	indirect pathways. Details are given in Section \ref{sec:extension} of the supplementary material.}
% \begin{figure}[t]
	%     \resizebox{0.35\columnwidth}{!}{% 
		%             \centering 
		%             \begin{tikzpicture}[node distance=1.1cm] 
			%                 \node[circle, draw=black] (U) {$A$}; 
			%                 \node[rectangle, below right=of U, draw=black, fill=gray!15] (G) {$G=(S_0,S_1)$};
			%                 \node[circle, below left=of U, draw=black] (Z) {$Z$}; 
			%                 \node[circle, above right=of G, draw=black] (W) {$R$};
			%                 \node[circle, below  =of Z, draw=black] (S) {$S$};
			%                 \node[circle, below  =of G, draw=black] (Y) {$Y$};
			%                 % Draw arrows with stealth arrow tips
			%                 \draw[->, -stealth] (U) -- (W);
			%                 \draw[->, -stealth] (U) -- (G); 
			%                 \draw[->, -stealth] (U) -- (Z); 
			%                 \draw[->, -stealth] (W) -- (G);
			%                 \draw[->, -stealth] (Z) -- (Y);
			%                 \draw[->, -stealth] (G) -- (Y);
			%                 \draw[->, -stealth] (Z) -- (S);
			%                 \draw[->, -stealth] (G) -- (S);
			%                 \draw[->, -stealth] (S) -- (Y);
			%                \draw[->, -stealth, bend right=20] (Y) to (W); % Add bend right here
			%                 % Draw the vertical edges of the trapezoid 
			%             \end{tikzpicture}  
		%     }
	%     \caption{ } 
	%     \label{fig:assumpts-illustration}
	% \end{figure}
\begin{figure}
	\centering
	\begin{tikzpicture}[node distance=1.1cm]
		\node[ ] (U) {$ $}; 
		\node[circle, below right=of U, draw=black, fill=gray!15, thick] (G) {$G $};
		\node[circle, below left=of U, draw=black, thick] (Z) {$Z$}; 
		\node[circle, above right =of Z, draw=black, thick] (A) {$A $};
		\node[circle, below  =of Z, draw=black, thick] (S) {$S$};
		\node[circle, below  =of G, draw=black, thick] (Y) {$Y$};
		\node[circle,  right=of Y, draw=black, thick] (W) {$R$};
		% Draw arrows with stealth arrow tips 
		\draw[->, -stealth, thick] (A) -- (G); 
		\draw[->, -stealth, thick] (A) -- (Z);  
		\draw[->, -stealth, thick] (Z) -- (Y);
		\draw[->, -stealth, thick] (G) -- (Y); 
		\draw[->, -stealth, thick] (Z) -- (S);
		\draw[->, -stealth, thick] (G) -- (S);
		\draw[->, -stealth, thick] (S) -- (Y);
		\draw[->, -stealth, thick] (G) -- (W);
		\draw[->, -stealth, thick ] (Y) to (W); % Add bend right here
		\draw[->, -stealth, thick, bend left=45] (A) to (W); % Add bend right here
		% Draw the vertical edges of the trapezoid 
	\end{tikzpicture}
	\caption{The node $A$  represents the proxy variable, the node  $Z$ represents treatment variable, the node $G$ represents principal stratiﬁcation (latent variable), the node $S$ represents  survival status,  the node $R$ represents missingness indicator, and the node  $Y$ represents the outcome. We omit the observed covariates $C$ for simplicity.}
	\label{fig:causal}
\end{figure}

\begin{theorem}
	Under Assumptions \ref{assu:igno}-\ref{assu:proxy},  
	$\Delta_{\LL}$ is identifiable.
	\label{thm:iden}
\end{theorem}

Theorem \ref{thm:iden}  states that the causal parameter  $\Delta_{\LL}$ is identifiable by introducing a proxy variable that affects the outcome only through the principal strata, while requiring that the treatment variable does not directly affect the missingness mechanism.    The proof of this theorem is given in the supplementary material.  

\section{Estimation}
\label{Sec:Methodology} 
\label{sec:est}
The nonparametric identification results in the previous section provide useful insights. 
However, fully nonparametric estimation methods, such as sieve-based approaches \citep{li2023non}, while theoretically appealing, often become impractical when the number of covariates is moderate or large due to the curse of dimensionality.
In this section, we focus on estimating the parameter $\Delta_{\LL}$ using parametric  methods.
% Because 
% \begin{equation}
	% 	\begin{aligned}
		% 		&E\{Y(z) \mid G=\LL\}=\frac{E\{E(Y\mid Z = z, G=\LL,C) \pr(G = \LL\mid A,C)\}}{E\{\pr(G = \LL\mid A,C)\}},
		% 	\end{aligned}
	% 	\label{eq:estimation}
	% \end{equation}
% we need to build models to estimate $E(Y\mid Z = z,G =\LL, C)$ and $\pr(G = \LL\mid A,C)$ at first. Proof of Eq.(\ref{eq:estimation}) is provided in the Supplementary Material.
Our estimation procedure will proceed in three steps, which is parallel with the three key assumptions \ref{assu:mono}-\ref{assu:proxy} of the previous section.

First, we impose two parametric models for estimating the proportions of principal strata under Assumption \ref{assu:mono}. Specifically,  we consider,
\begin{gather*}
	s_1(A,C;\beta_1)=    \pr(S=1 \mid Z =1,A,C),\\
	s_{0/1}(A,C;\beta_2)=   {\pr(S=1 \mid Z=0, A,C)}/{\pr(S=1 \mid Z=1, A,C)} ,
\end{gather*} 
where $s_1(A,C;\beta_1)$ and $s_{0/1}(A,C;\beta_2)$ are two parametric models bounded between 0 and 1, while $\beta_1$ and $ \beta_2$ are two unknown parameters to be estimated. {The rationale for the above modeling method is to ensure monotonicity assumption \ref{assu:mono}, particularly for the model $ s_{0/1}(A,C;\beta_2) $. Specifically, the parametric model $ s_1(A,C;\beta_1) $ represents the conditional probability $ \pr(S_1=1 \mid A, C) $, which is naturally bounded between 0 and 1. Additionally, the parametric model $ s_{0/1}(A,C;\beta_2) $ captures the relative ratio $ \pr(S_0 = 1 \mid A, C) / \pr(S_1 = 1 \mid A, C) $, which is also bounded between 0 and 1 under monotonicity assumption. This holds because the relative ratio can be expressed as:  
	$$
	s_{0/1}(A,C;\beta_2)=\dfrac{\pr(S_0 = 1 \mid A, C)}{\pr(S_1 = 1 \mid A, C)} = \dfrac{\pr(G=\AT \mid A, C)}{\pr(G=\AT \mid A, C) + \pr(G=\CO \mid A, C)}. 
	$$}{A similar approach to modeling survival mechanisms is found in \cite{wang2017identification}.}  In practice, $s_1(A,C;\beta_1)$ and $s_{0/1}(A,C;\beta_2)$ can be specified as logistic or probit regression models. We use the maximum likelihood estimation method to obtain estimators  $\widehat\beta_1$ and $\widehat\beta_2$. The proportions of principal strata can be estimated as follows:
\begin{equation*}
	\label{eq:mon-est}
	\begin{gathered}
		{\widehat  \pr}(G = \DD\mid X)  = 1- s_1(A,C;\widehat \beta_1),~~
		{\widehat  \pr}(G = \LL\mid X)    = s_1(A,C;\widehat \beta_1)s_{0/1}(A,C;\widehat \beta_2),\\
		{\widehat  \pr}(G = \LD\mid X)  =1 - \widehat  \pr(G = \DD\mid X)-  \widehat \pr(G = \LL\mid X).
	\end{gathered}
\end{equation*}
%  is a known parametric model, and $\beta_2$ is the parameters to estimate. 

Next, we consider applying a parametric model to estimate the missingness mechanism under Assumption \ref{assu:shadow}. Specifically, we develop a parametric model for the conditional probability $\mathrm{pr}(R = 1 \mid S = 1, A, C, Y)$, denoted as $m_1(A, C, Y; \alpha)$, where $\alpha$  is an unknown parameter to be estimated. We obtain an estimator of $\alpha$  through  solving the following estimating equation:
\begin{align}
	\label{eq:missing-model}
	\mathbb{P}_n\bigg[\bigg\{\cfrac{R}{m_1(A,C,Y;\alpha)}-1\bigg\}Sh_1(A,C,Z)\bigg] = 0,
\end{align}
where $\mathbb{P}_n(V)=\sum_{i=1}^n V_i / n$ for a generic variable $V$, and $h_1(A,C,Z)$  
is an arbitrary vector of functions of  $(A,C,Z)$   with dimension no smaller than that of
$\alpha$.

Finally, we apply parametric methods to fit the outcome model under Assumption \ref{assu:proxy}. Specifically, we consider two parametric models: $\mu_{1g}(C; \gamma_{1g})$ for the conditional probability $\pr(Y = 1 \mid Z = 1, G = g, C)$ and $\mu_{0g}(A, C; \gamma_{0g})$ for the conditional probability $\pr(Y = 1 \mid Z = 0, G = g, A, C)$.  For the binary outcome $Y$, the  model can be specified as a logistic or probit regression model. 
We consider the following estimating equations for estimating $\gamma_{{zg}}$:
\begin{gather*}
	\mathbb{P}_n\Bigg[\dfrac{RZS}{{m_1(A,C,Y;\widehat\alpha)}}\Bigg\{Y -\textstyle\sum_{g\in\{\LL,\LD\}}\mu_{1g}(C;\gamma_{1g}){\dfrac{\widehat{\pr} (G=g\mid A,C)}{s_1(A,C;\widehat\beta_1)}}\Bigg\}h_2(A,C) \Bigg]= 0,\\        \mathbb{P}_n\bigg[\dfrac{R(1-Z)S}{m_1(A,C,Y;\widehat\alpha)} \big\{Y-\mu_{0\LL}(A,C;\gamma_{0\LL})\big\}h_3(A,C)\bigg]
	=0,
	% \addlinespace[0.5mm]
\end{gather*}
where $ h_2(\cdot) $ and $ h_3(\cdot) $ are arbitrary functions of $ (A,C) $, with $ h_2(\cdot) $ requiring its dimension to be no less than the sum of the dimensions of $ \gamma_{1\LD} $ and $ \gamma_{1\LL} $, and $ h_3(\cdot) $ requiring its dimension to be no less than that of $ \gamma_{0\LL} $.

% is an arbitrary function of $(A,C)$ that requires its dimension to be no less than the sum of the dimensions of $\gamma_{1\LD}$ and $\gamma_{1\LL}$, while $h_3(\cdot)$ is an arbitrary function of $(A,C)$ that requires its dimension to be no less than that of $\gamma_{0\LL}$. 
%

Based on the above estimation results and the expression in \eqref{eqn:deltass}, we can estimate the causal parameter $\Delta_{\LL}$ using the following equation: 
\begin{equation*}
	\begin{aligned}
		&\widehat\Delta_{\LL}=\frac{\mathbb{P}_n\{\mu_{1\LL}(C;\widehat\gamma_{1\LL}) \widehat\pr(G = \LL\mid A,C)\}}{\mathbb{P}_n\{\widehat\pr(G = \LL\mid A,C)\}}-\frac{\mathbb{P}_n\{\mu_{0\LL}(A,C;\widehat\gamma_{0\LL}) \widehat\pr(G = \LL\mid A,C)\}}{\mathbb{P}_n\{\widehat\pr(G = \LL\mid A,C)\}}.
	\end{aligned}
	\label{eq:estimation}
\end{equation*}
Following the standard asymptotic theory \citep{tsiatis2006semiparametric}, the estimator $\widehat\Delta_{\LL}$ is asymptotically normal. {The specific variance expression and the proof of its asymptotic normality can be found in Section \ref{sec:proof-asym} of the supplementary material.} In practice, the asymptotic variance can be obtained using the bootstrap method.

{We also note that the estimation method proposed in this section may depend on the selection of certain auxiliary functions  $ h_j(\cdot) $ ($ j = 1, 2, 3 $). 
	% Although this selection does not affect consistency, these functions play a critical role in achieving efficiency. 
	Following the framework introduced by  Newey and McFadden (1994) \citep{newey1994large},  the optimal choice of auxiliary functions should be derived as the conditional expectations of the partial derivative of the relevant moment restrictions  \citep{BowdenSIM2011,clarke2015estimating}.   
	For instance, in the estimating equation for the parameter  $ \gamma_{0\LL} $, to minimize the estimation  variance, the optimal choice of $ h_3(\cdot) $ is: 
	$$
	h_3^{\text{eff}}(A,C) = \bigg(\E\bigg[\dfrac{\partial}{\partial \gamma_{0\LL}} \dfrac{R(1-Z)S}{m_1(A,C,Y; \alpha_0)} \big\{Y - \mu_{0\LL}(A,C;\gamma_{0\LL})\big\} \bigg| A,C\bigg]\bigg)^\T.
	$$
	However, it should be noted that in our case, where \( Y \) is binary, deriving a closed-form expression for  the optimal choice \( h_3^{\text{eff}}(A,C) \) is difficult. In practice, using a relatively simple form for 
	$h_3(\cdot)$ preserves the consistency of the estimator, although it may incur some efficiency loss. Therefore, adopting basic specifications such as \( h_3(A,C) = (1,A,C)^\T \)  often provides a favorable trade-off between computational tractability and finite-sample performance, as demonstrated in our simulation studies.
}
\section{Bounds}
\label{sec:bounds}
When the identification assumptions are violated, the estimation of bounds of principal causal effects may be preferred.  In this section, we consider the partial identifiability of $\Delta_{\LL}$ under Assumptions~\ref{assu:igno} and \ref{assu:mono}. Let $\theta_{z s t}(X)= \operatorname{pr}\{Y(z)=1 \mid S(0)=s, S(1)=t,X\}$,   $\pi_z(X)=\operatorname{pr}\{Y(z)=1 \mid S(z)=1,X\}$, and   $\varphi(X)=\pr(G=\LL\mid X)/\pr(G=\LL)$,  such that $\Delta_{\LL}=\E[\{\theta_{111}(X)-\theta_{011}(X)\}\varphi(X)]$.  
Let    $\gamma(X)=\operatorname{pr}\{S(0)=1 \mid S(1)=1,X\}$.   Following \cite{hudgens2006causal} and \cite{long2013sharpening}, we  note that, 
\begin{align*} 
	\pi_1(X)&  =\operatorname{pr}\{Y(1)=1 \mid S(0)=0, S(1)=1,X\} \operatorname{pr}\{S(0)=0 \mid S(1)=1,X\} \\
	&\quad+\operatorname{pr}\{Y(1)=1 \mid S(0)=1, S(1)=1,X\} \operatorname{pr}\{S(0)=1 \mid S(1)=1,X\} ,
\end{align*}
that is, $\pi_1(X)=\gamma(X) \theta_{111}(X)+\{1-\gamma(X)\} \theta_{101}(X) $, which further implies that 
\begin{align}
	\label{eq:ratio-form}
	\Delta_{\LL}= \E\bigg[\bigg\{\dfrac{\pi_1(X)}{\gamma(X)} - \dfrac{ 1 - \gamma(X)}{\gamma(X)} \theta_{101}(X)-\theta_{011}(X)\bigg\}\varphi(X)\bigg].
\end{align} 
The parameter $\gamma(X)$ and $\varphi(X)$ are identifiable  under Assumptions \ref{assu:igno} and \ref{assu:mono} through $$\gamma(X)=\dfrac{\mathrm{pr}(S = 1\mid Z=0,X)
}{ \pr(S=1\mid Z = 1, X)
} ~~\text{and}~~\varphi(X)=\dfrac{\mathrm{pr}(S = 1\mid Z=0,X)
}{ \E\{\pr(S=1\mid Z =0, X)\}
},$$ respectively, while the identification of $\theta_{101}(X)$ and $\theta_{011}(X)$ further requires   Assumptions \ref{assu:shadow} and \ref{assu:proxy}. Without these two assumptions, it is generally difficult to achieve nonparametric identification for $\theta_{101}(X)$ and $\theta_{011}(X)$. From   \eqref{eq:ratio-form}, it can be observed that $\gamma(X)$ serves as an adjustment factor for the unidentifiable terms   $\pi_1(X)$ and $\theta_{101}(X)$.  Specifically, when $\gamma(X)=1$, implying no compliers given the covariates $X$, the non-identifiability issues of $\theta_{101}(X)$ may not need to be addressed, and the impact of uncertainty from  $\pi_1(X)$ is also minimized.

Next, we explore the upper and lower bounds of each quantity separately  under Assumption \ref{assu:igno}. Define $\xi_{zr}(X)=\pr(Y  =1\mid Z=z,S  =1,R =r,X)$, where $\xi_{z1}(X)$ is identifiable, but $\xi_{z0}(X)$ is not. Let $\delta_{z}(X) $ denote  the conditional probability $\delta_{z}(X)=\pr(R=1\mid Z=z,S=1,X)$, which can be identified  from the observed data. Note that under Assumption~\ref{assu:igno}, $\pi_1(X)$ can also be expressed as:
\begin{align*} 
	\pi_1(X)&  =\operatorname{pr}(Y =1 \mid  Z=1,S =1,R =1,X) \operatorname{pr}(R=1 \mid Z=1,S=1,X) \\
	&\quad+\operatorname{pr}(Y =1 \mid  Z=1,S =1,R =0,X)\operatorname{pr}(R =0\mid Z=1,S=1,X).
\end{align*}
Thus, we express the quantity $\pi_1(X)$ as $\pi_1(X)=\delta_{1}(X) \xi_{11}(X)+\{1-\delta_{1}(X)\} \xi_{10}(X)$. It is evident that by utilizing the range $[0,1]$ of $\xi_{10}(X)$, we can derive both the upper bound $\pi_1^u(X)=\delta_{1}(X) \xi_{11}(X) +1-\delta_{1}(X)$ and the lower bound $  \pi_1^l(X)=\delta_{1}(X) \xi_{11}(X)$ for $\pi_1(X)$.  In our motivating example in Section \ref{sec:application}, 
if all missing CD4 level changes are actually positive for treatment group individuals who survive after two years, then $\xi_{10}(X) = 1$ and the parameter $\pi_1(X)$ can reach its upper bound; if all missing changes are indeed negative, then $\xi_{10}(X) = 0$ and $\pi_1(X)$ can reach its lower bound.

%	
%	if the changes in CD4 level were all positive for every individual in the treatment group, who is ``survival" (see Section \ref{sec:application} for specific definition of ``survival") after two years but with missing data, then $\xi_{10}(X) = 1$, and $\pi_1$ can attain its upper bound; if the changes were all negative, then $\xi_{10}(X) = 0$, and $\pi_1$ can attain its lower bound.

Under Assumption \ref{assu:igno}, the quantity $\theta_{101}(X)$ denotes the conditional  probability  $\pr(Y=1\mid Z=z,G=\LD,X)$. In the absence of any domain knowledge, the sharp bound of  $\theta_{101}(X)$ should be $[0,1]$. In our motivating example, when the CD4 level changes under treatment of all always-survivors are negative,  $\theta_{101}(X)$ attains its lower bound, namely $\theta_{101}^l(X)=0$; if the changes are all positive, $\theta_{101}(X)$ attains its upper bound, namely $\theta_{101}^u(X)=1$. Under Assumption \ref{assu:mono},  the quantity $\theta_{011}(X)$ can be simplified as $ \mathrm{pr}\{Y(0) = 1\mid S(0) =1,X\}$. Similarly,  $\theta_{011}(X)$  can be expressed as
\begin{align*} 
	\theta_{011}(X)&  =\operatorname{pr}(Y =1 \mid  Z=0,S =1,R=1,X)\operatorname{pr}(R=1 \mid Z=0,S=1,X) \\
	&\quad+\operatorname{pr}(Y =1 \mid  Z=0,S =1,R=0,X) \operatorname{pr}(R=0 \mid Z=0,S=1,X) .
\end{align*}
Therefore, we have $\theta_{011}(X) =\delta_{0}(X) \xi_{01}(X)+\{1-\delta_{0}(X)\} \xi_{00}(X)$ under Assumption \ref{assu:igno}. By using the range $[0,1]$ of $\xi_{00}(X)$, we can obtain the upper bound $\theta_{011}^u(X)=\delta_{0}(X) \xi_{01}(X) +1-\delta_{0}(X)$ and the lower bound $\theta_{011}^l(X)=\delta_{0}(X) \xi_{01}(X)$ for $\theta_{011}(X)$.
In our motivating example, 
if all missing CD4 level changes are actually positive for control group individuals who  survive, then $\xi_{00}(X) = 1$ and $\theta_{011}(X)$ can reach its upper bound; if all missing changes are actually negative, then $\xi_{00}(X) = 0$ and $\theta_{011}(X)$ can reach its lower bound.
% {\red In our motivating example, if the changes of CD4 level were all positive for every ``survival" individual with missing data in the control group, then $\xi_{00}(X) = 1$, and $\theta_{011}$ can attain its upper bound; if the changes were all negative, then $\xi_{00}(X) = 0$, and $\theta_{011}$ can attain its lower bound.}
Note from the decomposition of $\pi_1(X)$ and $\theta_{011}(X)$ that if $\delta_{0}(X)$ and $\delta_{1}(X)$ approach 1, meaning that the missing probabilities of the survival in the treatment and  control groups are small, then the uncertainty introduced by the non-identifiability issues of $\xi_{00}(X)$ and $\xi_{10}(X)$ can be mitigated, leading to narrower bounds for $\pi_1(X)$ and $\theta_{011}(X)$, which aligns well with our intuition.
% Besides, if the missing mechanism is MCAR or MAR, 
% which means $\xi_{z1}(X) = \xi_{z0}(X)$ under assumption \ref{assu:igno}, then $\pi_1(X)$ and $\theta_{011}(X)$ can be identified as $\xi_{11}(X)$ and $\xi_{01}(X)$, respectively.

Combining all the previous results, we can derive the identifiable lower and upper bounds for  $  \Delta_{\LL}$  as  $  [\Delta_{\LL}^l,\Delta_{\LL}^u]$, where  \begin{align}
	\label{eq:ratio-form-2} 
	\begin{aligned}
		%          \Delta_{\LL}^l&=\max\left[ \E\left\{\dfrac{\pi_1^l(X)}{\gamma(X)} - \dfrac{ 1 - \gamma(X)}{\gamma(X)} \theta_{101}^u(X)-\theta_{000}^u(X)\right\},-1\right],\\
		\Delta_{\LL}^l&=\E\bigg[\max\bigg\{0,\dfrac{\pi_1^l(X)}{\gamma(X)} - \dfrac{ 1 - \gamma(X)}{\gamma(X)}\theta_{101}^u(X) \bigg\}\varphi(X)\bigg]-\E\big\{\theta_{011}^u(X)\varphi(X)\big\},\\
		\Delta_{\LL}^u&= \E\bigg[\min\bigg\{1,\dfrac{\pi_1^u(X)}{\gamma(X)}- \dfrac{ 1 - \gamma(X)}{\gamma(X)}\theta_{101}^l(X) \bigg\}\varphi(X)\bigg]-\E\big\{\theta_{011}^l(X)\varphi(X)\big\}. 
	\end{aligned}
\end{align} 
\begin{theorem}
	Under Assumptions \ref{assu:igno}-\ref{assu:mono},  $  [\Delta_{\LL}^l,\Delta_{\LL}^u]$ are the  {sharp}   bounds of   $\Delta_{\LL}$. 
	\label{thm:bound2}
\end{theorem}

{Theorem \ref{thm:bound2} indicates that the lower bound $ \Delta_{\LL}^l $ and the upper bound $ \Delta_{\LL}^u $ are sharp, meaning there exist some  distributions that can achieve these bounds. For example, as demonstrated in Section \ref{sec:bound-proof} of the supplementary material, the lower bound  $ \Delta_{\LL}^l $  is feasible when $ \xi_{10}(x) = 0 $, $ \theta_{101}(x) = 1 $, and $ \xi_{00}(x) = 1 $ hold simultaneously for any $ X = x $. Further,}  Theorem \ref{thm:bound2} is not only applicable to observational studies when the ignorable treatment assignment assumption holds, but also to randomized experiments.  
Since under randomized experiments, we can also establish bounds for $\Delta_{\LL}$ without using any covariates \citep{luo2023causal}. A natural question is the comparative analysis between using unadjusted bounds and using adjusted bounds. We introduce more notation to consider unadjusted bounds under randomization.  Let $\theta_{z s t}= \operatorname{pr}\{Y(z)=1 \mid S(0)=s, S(1)=t\}$, $\gamma=\operatorname{pr}\{S(0)=1 \mid S(1)=1\}$, $\pi_z=\operatorname{pr}\{Y(z)=1 \mid S(z)=1\}$, $\delta_{z}=\pr(R =1\mid Z=z, S =1)$, $\xi_{zr}=\pr(Y =1\mid Z=z,S=1,R=r)$. 
Similarly, we have $\pi_1=\delta_{1} \xi_{11}+(1-\delta_1) \xi_{10}$ and $\theta_{011} =\delta_{0} \xi_{01}+(1-\delta_0) \xi_{00}$. %In randomized trials, $\delta_{z}$ can be identified through $\pr(R=1\mid Z=z,S=1)$, and  $\xi_{z1}$ is identifiable through $\pr(Y =1\mid Z=z,S =1,R =r)$.
Then the identifiable upper and lower bounds of $\pi_1$ are $\pi_1^u = \delta_{1} \xi_{11}+(1-\delta_1)$ and $\pi_1^l = \delta_{1} \xi_{11}$. The sharp bound of $\theta_{101}$ is still $[0,1]$, namely $\theta_{101}^l = 0$, $\theta_{101}^u = 1$. The identifiable upper and lower bounds of $\theta_{011}$ are $\theta_{011}^u = \delta_{0} \xi_{01}+(1-\delta_0)$ and $\theta_{011}^l = \delta_{0} \xi_{01}$. Then  the   unadjusted bound   for $\Delta_{ss}$ can be expressed as   $  [\Delta_{\LL}^{l*},\Delta_{\LL}^{u*}]$, where
\begin{align}
	\label{eq:unadj-bound}
	\Delta_{\LL}^{l*}= \max\bigg\{0,\dfrac{\pi_1^l}{\gamma} - \dfrac{ 1 - \gamma}{\gamma}\theta_{101}^u \bigg\}-\theta_{011}^u,~\Delta_{\LL}^{u*}=
	\min\bigg\{1,\dfrac{\pi_1^u}{\gamma}- \dfrac{ 1 - \gamma}{\gamma}\theta_{101}^l \bigg\}-\theta_{011}^l .
\end{align}
% \begin{corollary}
	% Under Assumptions \ref{assu:igno}-\ref{assu:mono},  if the treatment $Z$ is randomly assigned, then an unadjusted bound of $\Delta_{ss}$ is 
	% \begin{align}
		% \label{eq:unadj-bound}
		% 	\left[\max\left\{0,\dfrac{\pi_1^l}{\gamma} - \dfrac{ 1 - \gamma}{\gamma}\theta_{101}^u \right\}-\theta_{011}^u,
		% 	\min\left\{1,\dfrac{\pi_1^u}{\gamma}- \dfrac{ 1 - \gamma}{\gamma}\theta_{101}^l \right\}-\theta_{011}^l\right].
		% \end{align}
	% \end{corollary}
\begin{proposition}
	\label{prop:bound} Under  randomized trial and  Assumption \ref{assu:mono},
	% the   adjusted bound of $\Delta_{ss}$ should be narrower than     
	% the adjusted bound is narrower than the unadjusted bound. Specifically, 
	% \begin{align*}
		% 	\E\left[\max\left\{0,\dfrac{\pi_1^l(X)}{\gamma(X)} - \dfrac{ 1 - \gamma(X)}{\gamma(X)}\theta_{101}^u(X) \right\}\right]-\E\left[\theta_{011}^u(X)\right]\geq\max\left\{0,\dfrac{\pi_1^l}{\gamma} - \dfrac{ 1 - \gamma}{\gamma}\theta_{101}^u \right\}-\theta_{011}^u,\\
		% 	\E\left[\min\left\{1,\dfrac{\pi_1^u(X)}{\gamma(X)}- \dfrac{ 1 - \gamma(X)}{\gamma(X)}\theta_{101}^l(X) \right\}\right]-\E\left[\theta_{011}^l(X)\right]\leq	\min\left\{1,\dfrac{\pi_1^u}{\gamma}- \dfrac{ 1 - \gamma}{\gamma}\theta_{101}^l \right\}-\theta_{011}^l.
		% \end{align*}
	we have $  [\Delta_{\LL}^l,\Delta_{\LL}^u]\subset [\Delta_{\LL}^{l*},\Delta_{\LL}^{u*}]$.
	\label{prop:bound}
\end{proposition}

Proposition \ref{prop:bound} implies that the adjusted bound will not be wider than the unadjusted bound, which intuitively makes sense since incorporating more information from covariates yields a narrower bound \citep{luo2023causal,long2013sharpening}. The proofs of  Theorem \ref{thm:bound2} and Proposition \ref{prop:bound} are given in the supplementary material.

\section{Simulation Studies}
\label{sec:sim}
\subsection{Simulation settings}
\label{sec:sim-settings}
In this section, we conduct simulation studies to investigate the finite sample performance of the proposed estimators. We consider data-generating settings according to Figure \ref{fig:causal}.   We generate the baseline covariates $(A,  C)$ from a bivariate normal distribution with $ E(A)= E(C)=0$, $\mathrm{var}(A)=\mathrm{var}(  C)=1$ and   $\mathrm{cov}(A,  C)=0  $.  We generate  a binary treatment $Z$  from   a Bernoulli distribution with probability $0.6$; that is $Z\sim \text{Bern}(0.6)$.  We consider the following generating mechanism for principal strata: 
\begin{equation}
	\label{eq:prop-principal}
	\begin{gathered}
		\mathrm{pr}(G = \DD\mid A,C)  = 1 - \mathrm{expit}(0.55+0.25A+C),\\
		\mathrm{pr}(G = \LL\mid A,C)  =\mathrm{expit}(0.55+0.25A+C)\mathrm{expit}(0.45-0.5A+C),\\
		\mathrm{pr}(G=\LD\mid A,C)  =\mathrm{expit}(0.55+0.25A+C)\left\{1-\mathrm{expit}(0.45-0.5A+C)\right\},
	\end{gathered}
\end{equation}
% where $\theta_1(A, C) =\mathrm{expit}(0.55+0.25A+C) $
% \begin{align*}
	% 	\theta_1(A,C;\beta_1) = \cfrac{1}{1 + \exp\left\{\beta_1^\T(A^\T,C^\T)^\T\right\}},\\
	% 	\theta_{0/1}(A,C;\beta_2) = \cfrac{1}{1 + \exp\left\{\beta_2^\T(A^\T,C^\T)^\T\right\}},
	% \end{align*}
% where $\beta_1 = (0.55,0.25,1)^\T,\beta_2 = (0.45,-0.5,1)^\T$. 
% Therefore, the generating mechanism for different principal strata is:
% \begin{align*}
	% 	\mathrm{pr}(G = \LL\mid A,C) &= \theta_1(A,C;\beta_1)  \theta_{0/1}(A,C;\beta_2),\\
	% 	\mathrm{pr}(G=\LD\mid A,C) &= \theta_1(A,C;\beta_1)\left\{1-\theta_{0/1}(A,C;\beta_2)\right\},\\
	% 	\mathrm{pr}(G = \DD\mid A,C) &= 1 - \theta_1(A,C;\beta_1).
	% \end{align*}
where $\mathrm{expit}(x)=\mathrm{exp}(x)/\{1+\mathrm{exp}(x)\}$. 
The survival status $S$ is determined by treatment $Z$ and the principal stratification $G$. The outcome variable  $Y^*$ is generated from the following  Bernoulli distribution:
\begin{gather*}
	\pr(Y^* =1\mid Z=1,G = \LL,C)=\mathrm{expit}(0.9+0.3C),\\
	\pr(Y ^* =1\mid  Z=1,G = \LD,C)=\mathrm{expit}(0.5+0.4C),\\
	\pr(Y^* =1\mid Z=0,G = \LL,C)=\mathrm{expit}(-0.5+0.3C).
\end{gather*}
We generate the missingness indicator  $R$  from Bernoulli distribution 
with probability	$\pr(R=1\mid Z,A,C,Y^*,S=1)=\mathrm{expit}(1.5+0.5A+1.1Y^*)$. 
We generate the final outcome $Y$ as follows: If $S=1$ and $R=1$, then $Y=Y^\ast$;  If $S=1$ and $R=0$, then $Y$ is missing; In all other cases, $Y$ is set to $\mathrm{NA}$.  
The true value of $\Delta_{\LL}$ is   0.33.  

{Additionally, to investigate the estimation performance under the scenario of mixed continuous and binary covariates, we include an additional simulation setting in Section \ref{sec:cov-simu} of the supplementary material. Except for baseline covariates $ A $ and $ C $, the data generation mechanisms for other variables remain consistent with those described in this subsection. The estimation results are very similar to the conclusions presented in the main text. 
}  
\subsection{Point estimation}
\label{ssec:point-est}
To estimate $\Delta_{\LL}$, we consider three approaches for comparative analysis. The first approach involves a complete case analysis or naive estimation, which compares only the difference in the outcome means for individuals with $(Z=1,S=R=1)$ versus those with $(Z=0,S=R=1)$,  without considering truncation by death or MNAR issues. 
Secondly, we use the method proposed by \cite{wang2017identification} to examine the performance if the MNAR issues are  ignored.
% , although they differ from ours in terms of identification conditions and estimation techniques. We are interested in examining its performance when outcomes missing not at random are  ignored. 
Specifically, we treat $A$ as the substitutional variable, satisfying Assumptions 4 and 5 in \cite{wang2017identification}.    Finally, we consider the proposed estimation approach outlined in Section \ref{sec:est}. {In the latter two approaches, we use logistic regression models for both $ s_1(A, C; \beta_1) $ and $ s_{0/1}(A, C; \beta_2) $, which are consistent with the true data-generating models. This consistency arises because, based on the parameterization in \eqref{eq:prop-principal}, we derive that the forms of $ s_1(A, C; \beta_1) $ and $ s_{0/1}(A, C; \beta_2) $ naturally align with logistic models:  
	{\footnotesize
	$$
	\begin{gathered}
		s_1(A, C; \beta_1)  = \pr(S=1 \mid Z=1, A, C) = \mathrm{pr}(G = \LL \mid A,C) + \mathrm{pr}(G = \LD \mid A,C)  = \mathrm{expit}(0.55 + 0.25A + C), \\
		s_{0/1}(A, C; \beta_2)  = \frac{\pr(S=1 \mid Z=0, A, C)}{\pr(S=1 \mid Z=1, A, C)} = \frac{\mathrm{pr}(G = \LL \mid A,C)}{\mathrm{pr}(G = \LL \mid A,C) + \mathrm{pr}(G = \LD \mid A,C)}  = \mathrm{expit}(0.45 - 0.5A + C).
	\end{gathered}
	$$}
	 Additionally, the auxiliary functions used in the proposed method described in Section \ref{sec:est} are defined as $ h_1(\cdot) = (1, A, C,Z)^\T $, $ h_2(\cdot) = \{\widehat{\pr}(G = \AT \mid Z = S = 1, A, C), \widehat{\pr}(G = \AT \mid Z = S = 1, A, C)C, \widehat{\pr}(G = \CO \mid Z = S = 1, A, C), \widehat{\pr}(G = \CO \mid Z = S = 1, A, C)C \}^\T $, and $ h_3(\cdot) = (1,A,C)^\T $.}

Table~\ref{tab:simu} presents the bias, root mean squared error, and coverage probabilities of $95 \%$ confidence intervals averaged across 1000 replications with different sample sizes $n=500,2000,5000$. {The 95\% coverage probabilities are calculated using 200 bootstrap samples.}   {It can be observed that the proposed estimator exhibits negligible bias, and its RMSE decreases as the sample size increases.} The $95 \%$ coverage probabilities of the proposed estimators are close to the nominal level in all scenarios. These results confirm our previous theoretical conclusions.   The naive method and the method proposed by \cite{wang2017identification} yield inconsistent estimates, large variances, and insufficient coverages, suggesting that those methods are inappropriate to be directly applied to situations where the outcomes are missing not at random.
\begin{table}[t]
	\centering
	\caption{Bias ($\times 100$), root mean square error (RMSE) ($\times 100$), and   95\%  coverage probability ($\times 100$) of estimating $\Delta_{\LL}$ under different sample sizes.  }
	\label{tab:simu}
	\resizebox{0.849\textwidth}{!}{
		\begin{tabular}{ccccc} 
			\toprule
			Estimation Method                               & Sample Size & Bias ($\times 100$) & RMSE ($\times 100$) & 95\% Coverage Probability ($\times 100$)  \\ 
			\midrule
			\multirow{3}{*}{The naive estimator}            
			& 500        & $-3.5$               & 7.8                & 92.0                         \\
			& 2000        & $-3.2$               & 4.7                & 83.9                         \\
			& 5000        & $-3.1$               & 3.8                & 69.3                         \\ 
			\midrule
			\multirow{2}{*}{ The method proposed~}                            
			& 500        & 82.39              & 175.1              & 35.3                          \\
			
			\multirow{2}{*}{ in \cite{wang2017identification}}
			& 2000        & 80.63              & 81.02              & 2.2                          \\
			& 5000        & 81.01              & 81.16              & 0                            \\ 
			\midrule
			\multirow{2}{*}{The   proposed estimator} 
			& 500        & -0.57               & 17.09               & 95.5                         \\
			& 2000        & 0.20               & 9.01               & 94.1                         \\
			& 5000        & 0.39               & 6.44               & 94.6                         \\
			\bottomrule
		\end{tabular}
	}
\end{table}

% \begin{figure}[h]
	%     \centering 
	% 	\includegraphics[width=0.6\linewidth]{pictures/etaZ_results.pdf}
	% 	\caption{The  simulation performance with partial violation of Assumption \ref {assu:shadow}.}
	% 	\label{fig:bounds-2}
	% \end{figure}
\subsection{Sensitivity analysis}
\label{ssec:sensitivity-an}
{In this section, we conduct a sensitivity analysis to evaluate the performance of our proposed method under partial violations of the identification assumptions. Specifically,  we first include the covariate $ A $ in the outcome model and use the following settings:
	$$
	\begin{gathered}
		\pr(Y^* = 1 \mid Z = 1, G = \LL, C, A) = \mathrm{expit}(0.9 + 0.1\eta A + 0.3C),\\
		\pr(Y^* = 1 \mid Z = 1, G = \LD, C, A) = \mathrm{expit}(0.5 + 0.1\eta A + 0.4C),\\
		\pr(Y^* = 1 \mid Z = 0, G = \LL, C, A) = \mathrm{expit}(-0.5 + 0.1\eta A + 0.3C),
	\end{gathered}
	$$
	where $ \eta $ reflects the degree of violation of Assumption \ref{assu:proxy}. We also modify the model in Section \ref{sec:sim-settings} for the missingness indicator $ R $ as follows:
	\begin{equation*}
		\label{eq:missing-sen}
		\pr(R = 1 \mid Z, A, C, Y^*, S = 1) = \mathrm{expit}(1.5 + \eta Z + 0.5A + 1.1Y^*),
	\end{equation*}
	where $ \eta $ is also used to reflect the degree of violation of Assumption \ref{assu:shadow}.} 
The rest of the data-generating models remain unchanged as described in Section \ref{sec:sim-settings}.
% {\red , where we consider a sample size of 2000.} 
In particular, when $\eta \neq 0$, the causal parameter of interest $\Delta_\LL$ cannot be identified. To explore the performance of our method, we continue to use the proposed methods in Section \ref{Sec:Methodology} for estimation. It is important to note that our employed models are fully correctly specified only when $\eta = 0$.   Figure \ref{fig:etazaresults2} presents the point and interval estimates of $\Delta_\LL$, where $\eta$ ranges from -2 to 2.  {We find that when $\eta$ is very small, the results are relatively robust. As $\eta$ increases, the bias becomes larger, which aligns with our expectations.} In all cases, however, the 95\% confidence intervals cover the true value.
\begin{figure}[t]
	\centering
	\includegraphics[width=0.687\linewidth]{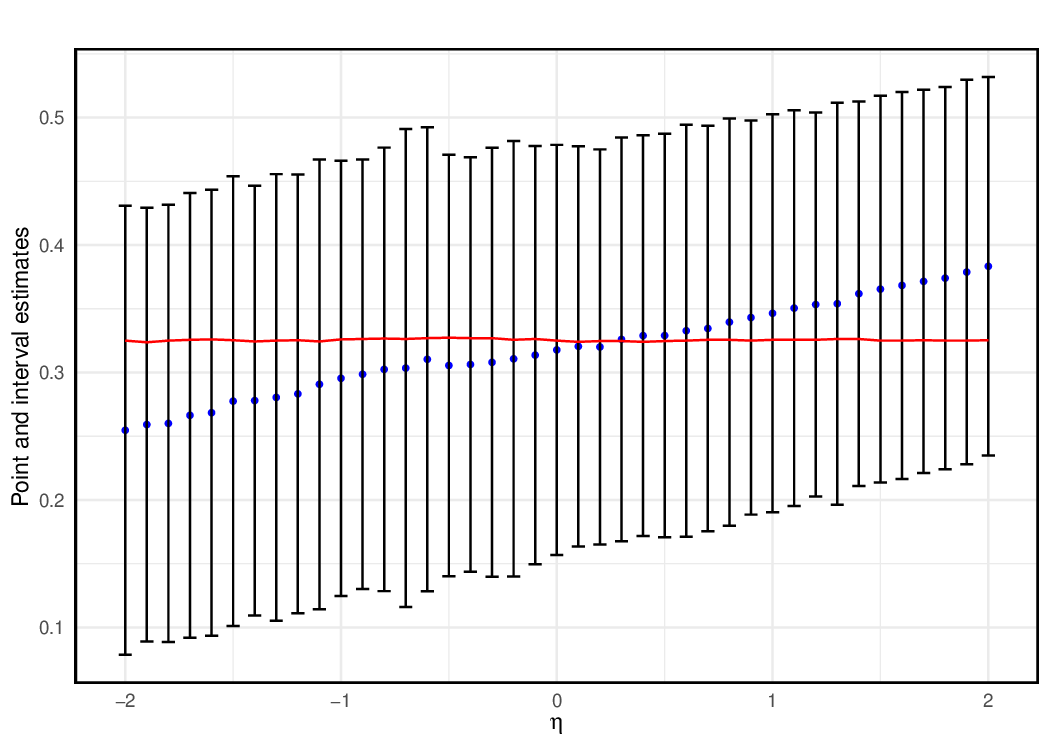}
	\caption{Sensitivity analysis when both Assumptions \ref{assu:shadow} and \ref{assu:proxy} are violated, with the red line representing the true causal effect.}
	\label{fig:etazaresults2}
\end{figure} 

\subsection{Bounds}
\label{ssec:bounds-simulation}
Besides point estimates and sensitivity analysis, we also consider the finite sample performance of the proposed bounds in Section \ref{sec:bounds} when both Assumptions \ref{assu:shadow} and \ref{assu:proxy} fail. {Speicifically,} {we generate the outcome that violates Assumption \ref{assu:proxy}:
	$$
	\begin{aligned}
		&\pr(Y^* = 1 \mid Z = 1, G = \LL, C, A) = \mathrm{expit}(0.9 + 0.02A + 0.3C), \\
		&\pr(Y^* = 1 \mid Z = 1, G = \LD, C, A) = \mathrm{expit}(0.5 + 0.02A + 0.4C), \\
		&\pr(Y^* = 1 \mid Z = 0, G = \LL, C, A) = \mathrm{expit}(-0.5 + 0.02A + 0.3C).
	\end{aligned}
	$$
	We violate Assumption \ref{assu:shadow} using the  missing data mechanism: $
	\pr(R = 1 \mid Z, A, C, Y^*, S = 1) = \mathrm{expit}(1.5 + 0.2 Z + 0.5A + 1.1Y^*).$}  The rest of the data-generating models remain unchanged as described in Section \ref{sec:sim-settings}.% {\red , where we consider a sample size of 8000.}  
%The results of these simulations are shown in Figure \ref{fig:bounds_comparison}.}
%Compared to Figure \ref{fig:bounds}, we observe the estimates for $ \Delta_{\AT} $ deviate significantly from the true value, and the estimates for the other bounds also display strange behavior. This highlights the sensitivity of the proposed method to violations of its underlying assumptions.}

\begin{figure}[ht] 
\centering
\includegraphics[width=0.6\linewidth]{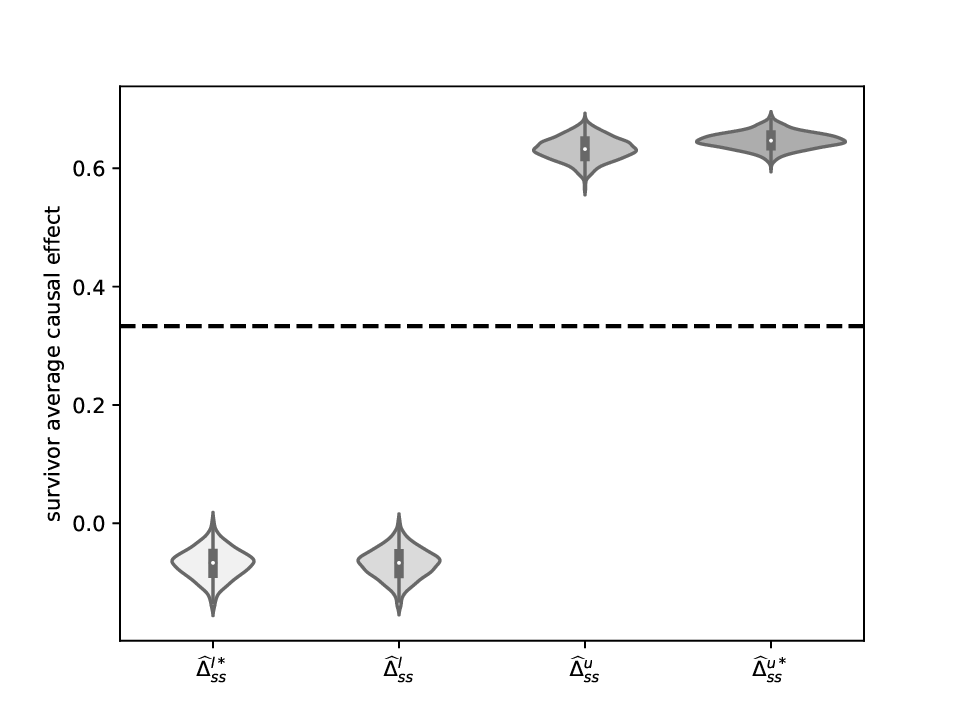}   
\caption{The violin plots of the estimated unadjusted lower bound $  \widehat\Delta_{\LL}^{l*  }$, the estimated adjusted lower bound $\widehat\Delta_{\LL}^{l}$, the estimated adjusted upper bound $\widehat\Delta_{\LL}^{u}$, and the estimated unadjusted upper bound $  \widehat\Delta_{\LL}^{u*}$,   where the dashed line indicates the true value 0.33.} 
\label{fig:bounds_comparison}
\end{figure}

{To avoid potential misspecification issues with parametric models for continuous covariates $ C $ and $ A $, which could bias the nonparametric bound estimates, we binarize these variables.   We achieve this by comparing each variable to its mean  and define $\tilde X=\{\mathbb{I}(C<0), \mathbb{I}(A<0)\}^\T$, where $\mathbb{I}(\cdot)$ is the indicator function. Since $ Z $ is randomized, Assumption \ref{assu:igno} holds for $Z\indep \{Y(1),Y(0),S(1),S(0),\tilde X\}$; the monotonicity assumption (Assumption \ref{assu:mono}) also holds in this simulation setting. Therefore, using $ \tilde{X} $ in the bound calculation from \eqref{eq:ratio-form-2} remains valid.} We then  estimate $\gamma(\tilde X)$, $\delta_0(\tilde X)$, $\delta_1(\tilde X)$,   $\xi_{01}(\tilde X)$, and $\xi_{11}(\tilde X)$ nonparametrically  and substitute them into \eqref{eq:ratio-form-2} to obtain the estimated adjusted bounds $[\widehat\Delta_{\LL}^{l },\widehat\Delta_{\LL}^{u }]$. The estimated unadjusted bounds  $[\widehat\Delta_{\LL}^{l*},\widehat\Delta_{\LL}^{u*}]$ are calucalted based on \eqref{eq:unadj-bound}. Based on  {1000}  replications,  {Figure \ref{fig:bounds_comparison} shows the violin plot of the estimated unadjusted lower bound $\widehat\Delta_{\LL}^{l * }$,} the estimated adjusted lower bound $\widehat\Delta_{\LL}^{l  }$, the point estimate $\widehat\Delta_{\LL}$, the estimated adjusted upper bound $\widehat\Delta_{\LL}^{u}$, and the estimated unadjusted upper bound $\widehat\Delta_{\LL}^{u*}$, where the dashed line indicates the true value. It can be observed that  the estimated causal effect fluctuates around the true causal effect. Our proposed bounds consistently encompass the estimated causal effect. Additionally, two violin plots for the adjusted bounds are visually narrower and closer to the true value than the violin plots for the unadjusted bounds, which validates  Proposition 
\ref{prop:bound}.

% {\red  }

% \begin{figure}[t]
% 	\centering
% 	\includegraphics[width=0.6\linewidth]{pictures/Bound_violin8.eps}
% 	\caption{The violin plots of the estimated unadjusted lower bound $  \widehat\Delta_{\LL}^{l*  }$, the estimated adjusted lower bound $\widehat\Delta_{\LL}^{l}$, the point estimate $\widehat\Delta_{\LL}$, the estimated adjusted upper bound $\widehat\Delta_{\LL}^{u}$, and the estimated unadjusted upper bound $  \widehat\Delta_{\LL}^{u*}$, where the dashed line indicates the true value 0.33.}
% 	\label{fig:bounds}
% \end{figure}
\section{Application to HIV Data}
\label{sec:application}
% In this section, we apply our proposed method to the HIV dataset from the AIDS Clinical Trials Group (ACTG) Study 175, which was a randomized and double-blind trial \citep{hammer1996trial}. In this trial, 2476 HIV patients were randomly assigned to different treatment groups. After removing missing and abnormal values, we obtained {2139} samples, among which {532} were assigned to the group receiving zidovudine-only therapy, denoted as $Z=0$, and the remaining 1607 were assigned to the new therapy group, including zidovudine plus didanosine, zidovudine plus zalcitabine, and didanosine only, denoted as $Z=1$. We define $S=0$ to represent individuals who were censored, deceased, or experienced similarly adverse outcomes, including a decline in CD4 T cell count of at least 50, which is one of the events indicating progression to AIDS; whereas $S=1$ indicates the absence of the aforementioned events. We choose the outcome $Y$ as the difference between a patient's CD4 level at the end of 96 weeks and its baseline CD4 level, with $Y=1$ if the change is positive; otherwise, $Y=0$.   We define $R=0$ to represent individuals with missing outcomes and $R=1$ to indicate individuals without missing. Our analysis aims to compare the treatment effects of zidovudine-only therapy and new therapies in terms of CD4 cell level changes approximately 96 weeks after treatment initiation \citep{orkin2021long}.
\subsection{Descriptive analysis}
\label{ssec:des}
In this section, we apply our proposed method to the HIV dataset from the AIDS Clinical Trials Group (ACTG) Study 175, a randomized and double-blind trial \citep{hammer1996trial}. In this trial, the adults infected with HIV-I  were randomly assigned to different treatment groups. After removing the abnormal values, we obtained {2139} samples, of which {532} were assigned to the group receiving zidovudine-only therapy   (ZDV-only), denoted as $Z=0$, and the remaining {1607} were assigned to the new therapy group including zidovudine plus didanosine, zidovudine plus zalcitabine, and didanosine only, denoted as $Z=1$.  {This definition of treatment variable has been widely used in previous analyses based on ACTG study  \cite{tsiatis2008covariate,Lu2011SemiparametricEO,song2017partially}.} We define $S=0$ to represent individuals who were censored, died, or experienced severe adverse outcomes, including a decline in CD4 T cell count of at least 50, which is one of the events indicating progression to acquired immunodeficiency syndrome (AIDS); whereas $S=1$ indicates the absence of the above events. {We define the binary outcome $Y$ as an indicator of whether the change in CD4 count over 96 weeks is positive, where $
Y=1$ if the final CD4 count exceeds the baseline level, and 
$Y=0$ otherwise. This corresponds to using zero as the threshold for defining the outcome. We chose this threshold for two main reasons. First, from a clinical perspective, an increase in CD4 count is generally interpreted as a sign of immunological improvement in HIV patients, which aligns with the intended interpretation of a “positive” outcome. Second, defining the outcome in terms of net improvement provides a straightforward and interpretable criterion across subjects with different baseline levels. Thus, defining \(Y = 1\) when the CD4 count increases is both clinically meaningful and widely used in practice.
In Section~\ref{sec:thre} of the supplementary material,  we further assess the sensitivity of our findings to alternative threshold values used to define the outcome. The results remain consistent with those reported in the main text.}  We define $R=0$ to represent individuals with missing outcomes and $R=1$ to represent those without missing outcomes. {Our analysis relies on the no-interference assumption inherent in SUTVA. The original publication \citep{hammer1996trial} did not raise interference as a concern, suggesting that the trial investigators did not view it as a relevant issue in this context. Several aspects of the trial design further support this view.
First, ACTG175 is a randomized, double-blind, placebo-controlled trial, in which patients were assigned to treatment arms using a blocked randomization procedure. Randomization was implemented centrally and independently at each clinical site, minimizing the potential for coordination or information sharing among patients across treatment arms.
Second, participants were recruited from 52 geographically dispersed clinical sites across the United States and Puerto Rico, making direct interactions between patients at different sites unlikely.
Third, the trial was conducted under strict protocols, with no evidence suggesting that patient interactions could have led to spillover effects. Based on these considerations, we believe that the no-interference assumption is reasonable in this study setting.} 
Assumption \ref{assu:igno} is naturally satisfied for this randomized trial.
We aim to estimate the treatment effects of zidovudine-only therapy and new therapies on the changes of CD4 cell counts approximately 96 weeks after treatment initiation \citep{orkin2021long}.

\begin{table}[t]
\centering
\caption{Descriptive analysis for baseline covariates. Means and standard errors (in parenthesis) for different subpopulations are reported.} 
\resizebox{0.995\textwidth}{!}{  
\begin{tabular}{ccccccc}
	\toprule
	& $(Z = 1,S = 1 ,R=0)$ & $(Z = 1,S = 0,R=0 )$ & $(Z = 0,S=1,R=0) $ & $(Z=0,S=0,R=0) $ & $(Z = 1,S=1,R=1) $ & $(Z=0,S=1,R=1) $ \\ 
	\hline
	The number of units  &  332 &469  &    92 &  236& 806 & 204              \\
	\texttt{age} $(X_1)$     & 35.22 (8.05) & 34.98 (9.07) & 35.68 (9.10) & 35.41 (9.68) & 35.43 (8.67)       & 34.80 (7.68)     \\
	\texttt{wtkg} $(X_2)$    & 75.31 (13.81) &74.21 (13.50)  & 76.60 (13.87) & 77.43 (13.74) & 74.97 (12.90)      & 74.23 (12.12)    \\
	\texttt{gender} $(X_3)$  & 0.82 (0.39) & 0.84 (0.37) & 0.77 (0.42) & 0.85 (0.36) & 0.84 (0.37)        & 0.79 (0.41)      \\
	\texttt{homo} $(X_4)$    & 0.65 (0.48) & 0.66 (0.48) & 0.60 (0.49) & 0.66 (0.48)     & 0.68 (0.47)        & 0.64 (0.48)      \\
	\texttt{drugs} $(X_5)$   & 0.18 (0.38) & 0.13 (0.33) & 0.18 (0.39) &0.11 (0.31)    & 0.12 (0.33)        & 0.10 (0.30)      \\
	\texttt{str2} $(X_6)$    & 0.57 (0.50) & 0.64 (0.48) & 0.58 (0.50) &0.63 (0.48)    & 0.57 (0.50)        & 0.52 (0.50)      \\
	\texttt{symptom} $(X_7)$ & 0.12 (0.32) &0.24 (0.43) & 0.15 (0.36) &0.21 (0.41)   & 0.16 (0.37)        & 0.13 (0.33)      \\
	\texttt{karnof} $(X_8)$  & 95.24 (6.19) &94.43 (6.40) & 95.33 (5.83) &95.17 (6.15)   & 96.13 (5.31)       & 95.78 (5.86)     \\
	baseline CD4 level $(X_9)$                & 366.50 (122.02) & 317.22 (121.13) & 363.96 (110.10) & 324.45 (109.47)  & 361.49 (115.00)    & 381.62 (113.69)  \\
	$Y=1$                                     & NA               & NA               &          NA          &  NA                & 0.55 (0.50)        & 0.44 (0.50)      \\ \bottomrule
	\end{tabular}	}
	\label{tab:descrip}
\end{table}

The collected baseline covariates $X=(X_1,\ldots,X_9)$ include:  age   ($X_1$, \texttt{age}), weight at baseline    ($X_2$, \texttt{wtkg}), gender   ($X_3$,  
\texttt{gender}; 0 for female, 1 for male), homosexual activity indicator    ($X_4$, \texttt{homo}; 0 $ =$ no, 1 $  =$ yes),  intravenous drug use history  ($X_5$, \texttt{drugs}; 0 $ =$ no, 1 $ =$ yes),  antiretroviral history  ($X_6$,   \texttt{str2}; 0 for naive, 1 for experienced), symptomatic indicator   ($X_7$, \texttt{symptom}),  Karnofsky score ($X_8$, \texttt{karnof}; on a scale of 0-100)  and the baseline CD4 level ($X_9$).  Table \ref{tab:descrip} presents descriptive statistics for different subpopulations.  Except for the baseline CD4 level, all covariates performed similarly across subpopulations in terms of mean and standard error metrics.  Naively comparing the CD4 level change at approximately 96 weeks, as shown in the last row of Table \ref{tab:descrip}, indicates a significant improvement for the new therapies, with an estimated mean difference $  0.116$ and a 95\% confidence interval $ [0.038,0.191]$. However, as mentioned earlier, this estimate lacks a causal interpretation.

%$We analyze the dataset under the monotonicity assumption \ref{assu:mono}.  Specifically, the 
The descriptive analysis of the survival rate and the probability of observed outcomes among survivors in the HIV dataset can be calculated as follows. The overall survival rate $\pr(S=1)$ is estimated to be 67.0\%, with the probability of observed outcomes among survivors $\pr(R=1\mid S=1)$ estimated at 30\%. In the treatment group, the survival rate $\pr(S=1\mid Z=1)$ is estimated to be 70.8\%, and the probability of observed outcomes among survivors $\pr(R=1\mid Z=1,S=1)$ is estimated at 29.2\%. In the control group, the survival rate $\pr(S=1\mid Z=0)$ is estimated to be 55.6\%, with the probability of observed outcomes among survivors $\pr(R=1\mid Z=0,S=1)$ estimated at 31.1\%.  These findings empirically suggest Assumption \ref{assu:mono}.  Therefore, we conduct our subsequent analyses based on the monotonicity assumption.  We use the expressions in \eqref{eq:mon} to estimate the proportions of   principal strata. We find that 29.2\% of the patients will die regardless of the treatment assigned ($G=\DD$), 55.6\% of the patients will live regardless of the treatment assigned ($G=\LL$), and the remaining 15.2\% of the patients' survival status will be affected by the treatment assignment ($G=\LD$).  We focus on the average causal effect comparing the treatment level with the control within the always-survival group, that is, $\Delta_{\LL}$. 
\subsection{Analysis under Assumptions \ref{assu:shadow} and \ref{assu:proxy}}
We first conduct analysis under Assumptions \ref{assu:shadow} and \ref{assu:proxy}, and estimate $\Delta_{\LL}$ using the proposed method outlined in Section \ref{sec:est}. As discussed under Assumption \ref{assu:shadow}, we consider the treatment-independent missingness in this example, because the missingness of the CD4 level is more likely due to poor health condition rather than the treatment assignment.  
We select the baseline CD4 level $X_9$ as the proxy variable $A$ considered in Assumption \ref{assu:proxy}. 
As discussed below Assumption \ref{assu:proxy}, baseline CD4 levels reflect the overall health condition of HIV-I patients and can serve as a surrogate for the principal stratum \( G \). {Assumption \ref{assu:proxy}(i) suggests that once treatment \( Z \), principal strata \( G \), and covariates \( C \) are accounted for, baseline CD4 levels \( A \) do not directly affect changes in CD4 after treatment. This assumption is biologically reasonable, as CD4 recovery is primarily driven by the effectiveness of antiretroviral therapy  and the patient’s overall health rather than the initial CD4 count \citep{smith2004factors}. Antiretroviral therapy suppresses viral replication, enabling immune restoration regardless of baseline CD4 levels. The immune system regulates CD4 homeostasis dynamically, meaning that even individuals with low baseline CD4 can experience significant improvement if they respond well to treatment. Moreover, factors such as chronic inflammation and co-infections, captured by \( G \), play a major role in immune recovery. Studies have shown that CD4 trajectories vary widely among individuals with similar baseline CD4 levels, suggesting that immune recovery depends more on treatment response and overall health than on the initial CD4 count. Therefore, once key determinants of CD4 recovery are accounted for, it is reasonable to assume that baseline CD4 does not have a direct effect on CD4 changes.} Moreover, in terms of numerical validation, we can see in the second last row of Table \ref{tab:descrip} that $X_9$ exhibits significant variation in means across different strata, thus  supporting the plausibility of Assumption \ref{assu:proxy}(ii). 
The point estimate of $\Delta_{\LL}$ is $  0.151$, with a 95\% confidence interval $ [-0.086, 0.275]$. {The obtained confidence interval covers zero, indicating that the causal effect of $Z$ on $Y$ is not significant for the always-survivor group.}  For comparison, we also examine the estimation results when the missing data problem is ignored using the method proposed by Wang et al. (2017)\citep{wang2017identification}, where we use $A$ as the substitutions variable  satisfying their 
Assumptions 4 and 5.    The point estimate of their method is larger with the value 0.226, and the 95\% confidence interval is $  [0.119, 0.777]$. This result is statistically significant, similar to the naive result obtained by the last row of Table \ref{tab:descrip}. The likely reason for the discrepancy between the comparison method and ours is that individuals with poor health, such as those with low CD4 levels, may have dropped out of the study. The comparison method does not account for this dropout effect, leading to an upward bias in its estimate.

\subsection{Sensitivity analysis for Assumption \ref{assu:shadow} }
\label{ssec:sen-assumption3}
To evaluate the robustness of our proposed  estimator under deviations from Assumption~\ref{assu:shadow}, we conduct a sensitivity analysis parallel to Section~\ref{ssec:sensitivity-an}. Specifically, we extend the model for the missing data mechanism by introducing a direct effect of the treatment variable \(Z\), captured by a sensitivity parameter \(\eta\), thereby allowing for partial violations of Assumption~\ref{assu:shadow}. For any given sensitivity parameter \(\eta\), we adjust the original estimating procedure by modifying the estimating equation~\eqref{eq:missing-model} in Section~\ref{Sec:Methodology} as follows:
\[
\mathbb{P}_n\bigg[\bigg\{\cfrac{R}{m_1(Z,A,C,Y;\alpha,\eta)}-1\bigg\}Sh_1(A,C,Z)\bigg] = 0,
\]
where \(m_1(Z,A,C,Y;\alpha,\eta)=  \mathrm{expit}(\alpha_0 + \eta Z + \alpha_{1,a} A + \alpha_{1,c}^\T C + \alpha_{1,y}  Y)\) and \(\alpha=(\alpha_0 , \alpha_{1,a},\alpha_{1,c}^\T,\alpha_{1,y} )^\T\). Based on this adjusted formulation, we continue to apply the  estimation strategy outlined in Section~\ref{Sec:Methodology} to solve for the model parameters. %While this modification introduces a slight deviation from the original framework, the overall estimation approach remains consistent, allowing us to systematically evaluate the robustness of our results under varying degrees of deviation from the identification assumption.

\begin{figure}
\centering
\includegraphics[width=0.6\linewidth]{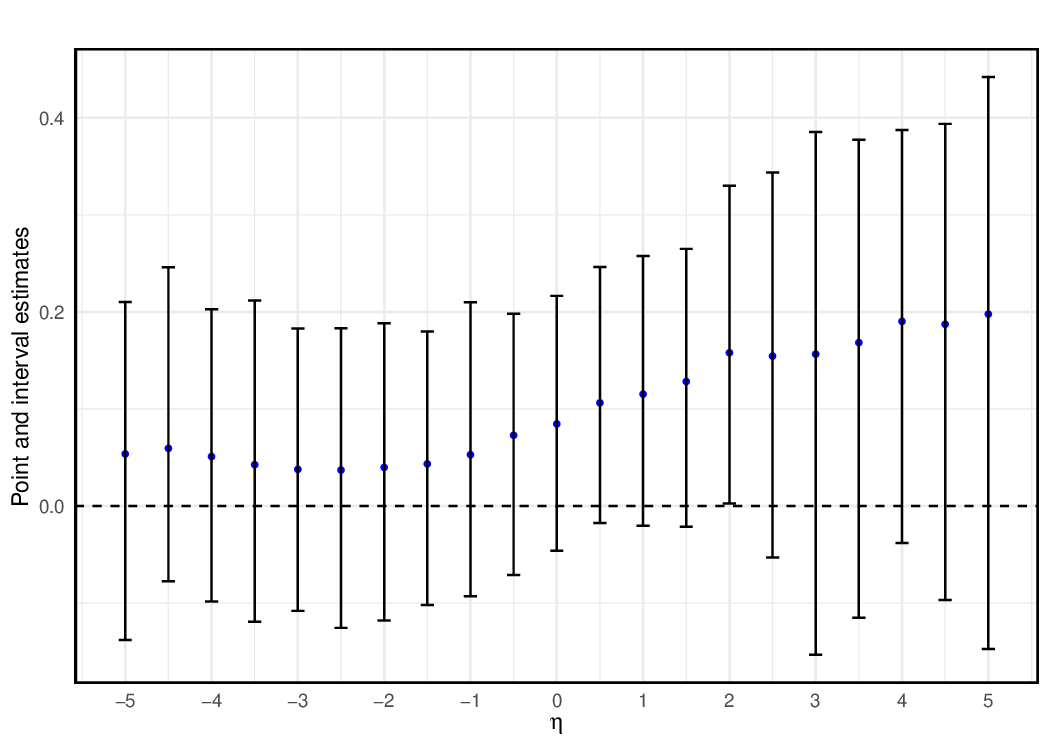}
\caption{Sensitivity analysis when   Assumption \ref{assu:shadow} is violated in the application.}
\label{fig:sensitivity-real}
\end{figure}

As shown in Figure~\ref{fig:sensitivity-real}, the point estimates of \(\Delta_{ss}\) are all larger than zero,  but the corresponding 95\% confidence intervals consistently include zero across the entire range of \(\eta\). This suggests that even under substantial deviations from the identification assumption \ref{assu:shadow}, the estimated effect \(\Delta_{ss}\) remains statistically insignificant, indicating strong robustness of the proposed method. These results imply that for individuals belonging to the principal stratum \(G =ss\), treatment assignment does not lead to significant changes in the outcome.

\subsection{Bounds analysis without Assumption \ref{assu:shadow} or \ref{assu:proxy}}\label{subsec:bounds}
% We note that the results obtained by Wang et al. (2017)\citep{wang2017identification} and the naive method are not consistent with the results obtained by our approach. 
{Next, we consider estimating the nonparametric bounds of $\Delta_{\LL}$ without imposing Assumption \ref{assu:shadow} or \ref{assu:proxy}, using the method proposed in Section \ref{sec:bounds}.}  For this randomized experiment, based on \eqref{eq:unadj-bound}, we initially obtain the unadjusted bounds $[ \Delta_{\LL}^{l*}, \Delta_{\LL}^{u*} ]$ for $\Delta_{\LL}$ as $ [-0.388, 0.567]$, where the 95\% confidence interval for $\Delta_{\LL}^{l*}$ is $  [-0.476, -0.311]$ and the 95\% confidence interval for $\Delta_{\LL}^{u*}$ is $  [0.474, 0.665]$. Additionally, after binarizing the baseline weight at the median, we calculate the adjusted bounds $[ \Delta_{\LL}^l, \Delta_{\LL}^u ]$ as $ [-0.173, 0.533]$ based on \eqref{eq:ratio-form-2}, where the 95\% confidence interval for $\Delta_{\LL}^l$ is $ [-0.489, -0.314]$ and the 95\% confidence interval for $\Delta_{\LL}^u$ is $ [0.460, 0.602]$. These two estimation results provide two important observations: first, they validate Proposition \ref{prop:bound}, indicating that any combination of covariates can further narrow the bounds in a randomized experiment. Second,  {both the adjusted and unadjusted bounds cover zero,} suggesting that the estimates of  $\Delta_{\LL}$  are not significant.   Furthermore, in this ACTG study, both the  ZDV-only and the new therapy are active treatments, and from a clinical point of view, the significant causal effects may not exist. All the above analyses indicate that new therapies may not significantly improve CD4 levels after 96 weeks among the always-survivor group \citep{trialists1999zidovudine,maenza1998combination}. This likely reflects that the ZDV-only regime is already effective for certain patients, and the addition of didanosine or zalcitabine in the new therapy group does not yield persistent or significant improvements in CD4 cell counts. As a result, no clear treatment advantage is observed for this stratum.

\section{Discussion}
\label{sec:dis}
In this paper, we investigate the identification, estimation, and bounds of the survivor average causal effect with outcomes that are both truncated by death and missing not at random. The problem considered in this paper is common in various fields, especially in clinical trials, yet it has not been sufficiently addressed before. Two key assumptions, Assumption \ref{assu:shadow} and Assumption \ref{assu:proxy}, are pivotal for nonparametric identification. The former requires that the treatment variable $Z$ has no direct effect on the missingness mechanism, while the latter necessitates the existence of a pretreatment covariate $A$ that does not directly affect the outcome variable. While both assumptions potentially hold  in our example, further discussions are needed in many practical problems. When these two crucial assumptions are violated, we also explore nonparametric bounds for the survivor average causal effect. 

{Although we binarize the covariates in Section \ref{ssec:bounds-simulation} of the simulation studies, the bounds proposed in Section \ref{sec:bounds} do not require the baseline covariates   to be discrete. The main reason for this binarization is to ensure the validity of the estimated bounds within a nonparametric framework. Specifically, if the covariates were not binarized, it would be necessary to introduce parametric models to handle continuous covariates. However, parametric models are susceptible to model misspecification, especially when multiple parametric models need to be specified. Such misspecification can lead to estimates that deviate from the true bounds, thereby compromising our goal of identifying the ranges of the parameters of interest under minimal modeling restrictions. Therefore, we choose to binarize the covariates to avoid imposing overly strict parametric assumptions and to facilitate a more straightforward nonparametric estimation process.}

{In practice, when the number of covariates is large, it is worth exploring LASSO-based variable selection or other modern machine learning techniques to improve model flexibility and performance.
However, standard LASSO methods are primarily designed for regression problems with a well-defined loss function, whereas our estimation approach relies on solving estimating equations specifically tailored for nonignorable missingness (e.g., estimating equation \eqref{eq:missing-model}).
Therefore, directly applying sparse penalization techniques such as LASSO is not straightforward in our setting.
Extending penalization strategies to estimating equation frameworks is an important and interesting direction for future research.} 

The proposed methods can {also} be improved or extended in various aspects. Firstly, our focus has been on identifying the survivor average causal effect when both the treatment and outcome variables are binary. It would be interesting to extend these results to continuous cases \citep{yang2019causal}. {Additionally, the no-interference assumption embedded in SUTVA may not always hold in medical studies, especially in  the context of communicable diseases. An important direction for future research is to explore violations of the SUTVA assumption in causal inference settings with outcomes truncated by death or missing not at random.}  Finally, Assumption \ref{assu:igno} assumes the absence of unmeasured confounders between the treatment and potential outcomes, which may be restrictive in some cases. Therefore, it would be of interest to identify the survivor average causal effect without Assumption \ref{assu:igno}\citep{luo2024causal,miao2018identifying,shi2020multiply}. The study of these issues is beyond the scope of this paper and we leave them as future research topics.

\bibliographystyle{apalike}
\bibliography{mybib}
\newpage
\renewcommand{\theproposition}{S\arabic{proposition}}
\renewcommand{\thetheorem}{S\arabic{theorem}}
\renewcommand{\theassumption}{S\arabic{assumption}}
\renewcommand{\thesection}{S\arabic{section}}
\renewcommand{\theequation}{S\arabic{equation}}
\renewcommand{\thelemma}{S\arabic{lemma}} 
\renewcommand{\thetable}{S\arabic{table}} 

{\centering \section*{Supplementray Material}}
In this supplementary material, we provide proofs of all theoretical results in
the main paper, additional simulation studies,  additional real data analyses, and an  extension of the theoretical framework with intermediate variables.

\setcounter{section}{0}

 	\section{Proof of Theorem \ref{thm:iden}}  
\label{sec:proof-thm1}
{\it Step} 1: we first prove  that
given Assumptions  \ref{assu:igno}-\ref{assu:proxy},  
the conditional distributions $  f(R=1\mid   S =1 ,A,C,Y)$ and $f(   Y\mid Z,S=1,A,C)$ are identifiable.
\label{th1}
%	\end{proposition}
\begin{proof}
\label{proof-thm1-1}
First, since $ (Z, A, C, S=1) $ is fully observed, the joint probability density function $ f(Z, S=1, A, C) $ is identifiable. Additionally, $ f(R=1, Y \mid Z, S=1, A, C) $ is also identifiable. Our focus is now on the identifiability of $ f(R=1 \mid S=1, A, C, Y) $.

We begin by noting the relationship:
$$
f(C, A, Z, S=1, Y, R) = f(C, A, Z, S=1) f(Y \mid C, A, Z, S=1) f(R \mid C, A, Z, S=1, Y).
$$
Given Assumption \ref{assu:shadow}(ii), this expression can also be rewritten as:
$$
f(C, A, Z, S=1, Y, R) = f(C, A, Z, S=1) f(Y \mid C, A, Z, S=1) f(R \mid C, A, S=1, Y).
$$
Dividing both sides by $ f(C, A, Z, S=1) f(R \mid C, A, S=1, Y) $, we obtain:
\begin{equation}\label{eqn:cond-Y}
	\cfrac{f(Y, R=1 \mid C, A, Z, S=1)}{f(R=1 \mid C, A, S=1, Y)} = f(Y \mid C, A, Z, S=1).
\end{equation}
Summing over $ y \in \{0, 1\} $, we get:
$$
\sum_{y=0,1} \cfrac{f(y, R=1 \mid C, A, Z, S=1)}{f(R=1 \mid C, A, S=1, y)} = \sum_{y=0,1} f(y \mid C, A, Z, S=1) = 1,
$$
which leads to the following matrix equation:
$$
\begin{pmatrix}
	f_{01 \mid 11}(C, A) & f_{11 \mid 11}(C, A) \\
	f_{01 \mid 01}(C, A) & f_{11 \mid 01}(C, A)
\end{pmatrix}
\begin{pmatrix}
	\cfrac{1}{f(R=1 \mid S=1, A, C, Y=0)} \\
	\cfrac{1}{f(R=1 \mid S=1, A, C, Y=1)}
\end{pmatrix}
=
\begin{pmatrix}
	1 \\
	1
\end{pmatrix},
$$
where $ f_{yr \mid zs}(C, A) $ denotes $ f(Y=y, R=r \mid C, A, Z=z, S=s) $.

To identify $ f(R=1 \mid S=1, A, C, Y) $, it suffices to show that the matrix:
$$
\Sigma := \begin{pmatrix}
	f_{01 \mid 11}(C, A) & f_{11 \mid 11}(C, A) \\
	f_{01 \mid 01}(C, A) & f_{11 \mid 01}(C, A)
\end{pmatrix}
$$
is of full rank. Since 
$  
f(R=1, Y \mid Z, S=1, A, C) = f(R=1 \mid C, A, S=1, Y) f(Y \mid C, A, Z, S=1),
$ 
it follows that the full rank of $ \Sigma $ is equivalent to the full rank of the matrix:
$$
\begin{pmatrix}
	f(Y=0 \mid C, A, Z=1, S=1) & f(Y=1 \mid C, A, Z=1, S=1) \\
	f(Y=0 \mid C, A, Z=0, S=1) & f(Y=1 \mid C, A, Z=0, S=1)
\end{pmatrix}.
$$
This matrix is of full rank under Assumption \ref{assu:shadow}(i). Hence, the identifiability of $ f(R=1 \mid S=1, A, C, Y) $ is established. According to \eqref{eqn:cond-Y}, $f(Y\mid Z,S=1,A,C)$ is also identifiable.

\end{proof}
%	\begin{corollary}
{\it Step} 2: We next claim that, given Assumptions \ref{assu:igno}-\ref{assu:proxy}, the following conditional distributions are identifiable:

\begin{enumerate}
	\item[(1)] {Missing data distribution:}  $
	f(R=0, Y \mid Z , S=1, A, C) $
	
	% \item[(2)] {Conditional distribution for survivors:} 
	%  $
	% f(Y \mid Z , S=1, A, C).
	% $ 
	
	\item[(2)] {Conditional distribution:} 
	$ 
	f(Y \mid Z=0, G=\LL, A, C) 
	$  is identified.
	\item[(3)]  {Conditional distribution:}   $ f(G=g \mid Z=S=1, A, C)$ is identified for $g\in\{\LL,\LD,\DD\}$.
	
	\item[(4)] {Expected value:} $ \E(Y \mid Z=1, G=\LL, C) $ is identified.
\end{enumerate}

\begin{proof}
	We proceed with the proof as follows:
	
	\begin{itemize}
		\item[] \textbf{Proof of (1):}  
		From the first step, $ f(Y \mid Z, S=1, A, C)$ is identifiable. Additionally, $ f(R=0 \mid A, S=1, C, Y) = 1 - f(R=1 \mid A, S=1, C, Y)$ is also identifiable. Therefore:
		$$
		f(R=0, Y \mid Z, S=1, A, C) = f(Y \mid Z, S=1, A, C) f(R=0 \mid A, S=1, C, Y)
		$$
		is identifiable.
		
		%  \item[] \textbf{Proof of (2):}  
		% From the first step, $ f(Y \mid Z, S=1, A, C)$ is identifiable.
		
		\item[] \textbf{Proof of (2):}  
		Under Assumption \ref{assu:mono}, we have:
		$$
		f(Y \mid Z=0, S=1, A, C) = f(Y \mid Z=0, G=\LL, A, C),
		$$
		which  implies that    $ 
		f(Y \mid Z=0, G=\LL, A, C) 
		$  is identifiable.
		
		\item[] \textbf{Proof of (3):}  
		The proportion $ f(G=\LL \mid Z=S=1, A, C)$ can be identified using:
		\begin{align*}
			f(G=\LL \mid Z=S=1, A, C) 
			&= \frac{f(G=\LL \mid Z=1, A, C)}{f(S=1 \mid Z=1, A, C)} \tag{$ G \indep Z \mid A, C$} \\
			&= \frac{f(G=\LL \mid Z=0, A, C)}{f(S=1 \mid Z=1, A, C)} \\
			&= \frac{f(S=1 \mid Z=0, A, C)}{f(S=1 \mid Z=1, A, C)}.
		\end{align*}
		Consequently, $ f(G=\LD \mid Z=S=1, A, C)$ is identified as:
		$$
		f(G=\LD \mid Z=S=1, A, C) = 1 - f(G=\LL \mid Z=S=1, A, C).
		$$
		\item[] \textbf{Proof of (4):}  
		Under Assumption \ref{assu:proxy}(i), by direct calculation, we have:  
		\begin{align*}
			\E(Y \mid Z=1, S=1, A=a_1, C) &= \E(Y \mid G=\LL, Z=1, C) f(G=\LL \mid Z=1, S=1, A=a_1, C) \\
			&\quad + \E(Y \mid G=\LD, Z=1, C) f(G=\LD \mid Z=1, S=1, A=a_1, C), \\
			\E(Y \mid Z=1, S=1, A=a_2, C) &= \E(Y \mid G=\LL, Z=1, C) f(G=\LL \mid Z=1, S=1, A=a_2, C) \\
			&\quad + \E(Y \mid G=\LD, Z=1, C) f(G=\LD \mid Z=1, S=1, A=a_2, C),
		\end{align*}
		where $ a_1 \neq a_2$.  
		
		Since the coefficient matrix:
		$$
		\begin{bmatrix}
			f(G=\LL \mid Z=1, S=1, A=a_1, C) & f(G=\LD \mid Z=1, S=1, A=a_1, C) \\
			f(G=\LL \mid Z=1, S=1, A=a_2, C) & f(G=\LD \mid Z=1, S=1, A=a_2, C)
		\end{bmatrix}
		$$
		is of full rank under Assumption \ref{assu:proxy}(ii), the identifiability of $ \E(Y \mid Z=1, G=\LL, C)$ can be achieved.
	\end{itemize}
\end{proof}

%	\begin{proposition}
	{\it Step} 3:
	We finally claim that under Assumptions \ref{assu:igno}-\ref{assu:proxy}, the conditional expectations $ \E(Y(0) \mid G=\LL)$ and $ \E(Y(1) \mid G=\LL)$ are identifiable.
	
	\begin{proof}
		We start by showing the identifiability of $ f(Y(0) \mid G=\LL, A, C)$. Note that:
		\begin{align*}
			f(Y \mid Z=0, G=\LL, A, C) 
			&= f(Y(0) \mid Z=0, G=\LL, A, C) \tag{Consistency} \\
			&= \frac{f(Y(0), G=\LL \mid Z=0, A, C)}{f(G=\LL \mid Z=0, A, C)} \\
			&= \frac{f(Y(0), G=\LL \mid A, C)}{f(G=\LL \mid A, C)} \tag{$ Z \indep \{Y(z), S(z)\} \mid A, C$} \\
			&= f(Y(0) \mid G=\LL, A, C).
		\end{align*}
		Thus, $ f(Y(0) \mid G=\LL, A, C)$ is identifiable.
		
		Next, we show the identifiability of $ f(Y(0) \mid G=\LL)$:
		\begin{align*}
			f(Y(0) \mid G=\LL) 
			&= \int f(Y(0), A, C \mid G=\LL) \, d\mu(A) \, d\mu(C) \\
			&= \int f(Y(0) \mid A, C, G=\LL) f(A, C \mid G=\LL) \, d\mu(A) \, d\mu(C).
		\end{align*}
		It suffices to identify $ f(A, C \mid G=\LL)$.
		
		Now, note that:
		$$
		f(A, C \mid G=\LL) = \frac{f(G=\LL \mid A, C) f(A, C)}{\int f(G=\LL \mid A, C) f(A, C) \, d\mu(A) \, d\mu(C)},
		$$
		which can be identified.
		
		Therefore, $ \E(Y(0) \mid G=\LL)$ is identifiable. Similarly, $ \E(Y(1) \mid G=\LL)$ can be identified by analogous steps.
		
	\end{proof}
		\section{Proof of Theorem \ref{thm:bound2}}  
		\label{sec:bound-proof}
		\begin{proof}
			To prove Theorem \ref{thm:bound2}, it is sufficient to prove (i) $\Delta_{\LL}^l\leq \Delta_{\LL}$ and (ii) $\Delta_{\LL}^u\geq \Delta_{\LL}$.
			From the definitions of $\pi_1^l(X),\pi_1(X),\pi_1^u(X),\theta_{101}^l(X),\theta_{101}(X),\theta_{101}^u(X),\theta_{011}^l(X),\theta_{011}(X),\theta_{011}^u(X)$, we know that  $\pi_1^l(X)\leq\pi_1(X)\leq\pi_1^u(X),\theta_{101}^l(X)\leq\theta_{101}(X)\leq\theta_{101}^u(X)$, and $\theta_{011}^l(X)\leq\theta_{011}(X)\leq\theta_{011}^u(X),a.s.$. Besides, $\theta_{111}(X)\in [0,1]$.
			Therefore,
			\begin{align*}
				\Delta_{\LL}&= \E\left[\left\{\theta_{111}(X)-\theta_{011}(X)\right\}\varphi(X)\right]\\
				&= \E\left[\left\{\dfrac{\pi_1(X)}{\gamma(X)} - \dfrac{ 1 - \gamma(X)}{\gamma(X)} \theta_{101}(X)-\theta_{011}(X)\right\}\varphi(X)\right]\\
				&= \E\left[\left\{\dfrac{\pi_1(X)}{\gamma(X)} - \dfrac{ 1 - \gamma(X)}{\gamma(X)} \theta_{101}(X)\right\}\varphi(X)\right] - \E\left[\theta_{011}(X)\varphi(X)\right]\\
				&\geq 
				\E\left[\max\left\{0,\dfrac{\pi_1^l(X)}{\gamma(X)} - \dfrac{ 1 - \gamma(X)}{\gamma(X)}\theta_{101}^u(X) \right\}\varphi(X)\right]-\E\left\{\theta_{011}^u(X)\varphi(X)\right\}\\
				&\triangleq\E\{\Delta_{\LL}^l(X)\}=\Delta_{\LL}^l,
			\end{align*}
			and 
			\begin{align*}
				\Delta_{\LL}&= \E\left[\left\{\theta_{111}(X)-\theta_{011}(X)\right\}\varphi(X)\right]\\
				&= \E\left[\left\{\dfrac{\pi_1(X)}{\gamma(X)} - \dfrac{ 1 - \gamma(X)}{\gamma(X)} \theta_{101}(X)\right\}\varphi(X)\right] - \E\left[\theta_{011}(X)\varphi(X)\right]\\
				&\leq 
				\E\left[\min\left\{1,\dfrac{\pi_1^u(X)}{\gamma(X)}- \dfrac{ 1 - \gamma(X)}{\gamma(X)}\theta_{101}^l(X) \right\}\varphi(X)\right]-\E\left\{\theta_{011}^l(X)\varphi(X)\right\}\\
				&\triangleq\E\{\Delta_{\LL}^u(X)\}=\Delta_{\LL}^u.
			\end{align*}
		\end{proof}

		Next, we prove that the proposed bound is sharp, that is, $\Delta_{\LL}^l= \Delta_{\LL}$ and  $\Delta_{\LL}^u= \Delta_{\LL}$ can be achieved in some cases.  The goal is to show that both the lower and upper bounds are feasible, meaning there exists some distribution that achieves them. We will only prove the feasibility of the lower bound, and the sharp upper bound follows similarly.
		
		\begin{proof} 
			To prove that $ \Delta_{\LL}^l $ is sharp, we aim to show that there exists a distribution that can indeed achieve this lower bound $ \Delta_{\LL}^l(x) $ for any $ X = x $. 
			By definition, we have the following expression for $\Delta_{\LL}(x)$:
			$$
			\Delta_{\LL}(x) = \{\theta_{111}(x) - \theta_{011}(x)\}\psi(x) = \bigg\{\frac{\pi_1(x)}{\gamma(x)} - \frac{1 - \gamma(x)}{\gamma(x)} \theta_{101}(x) - \theta_{011}(x)\bigg\}\psi(x),
			$$
			where $\gamma(x)$ and $\psi(x)$ are identifiable. 
			Using the notations given in the main text, we expand the above equation further:
			\begin{align*}
				\Delta_{\LL}(x) &=\bigg[ \frac{\delta_1(x) \xi_{11}(x) + \{1 - \delta_1(x)\} \xi_{10}(x)}{\gamma(x)} - \frac{1 - \gamma(x)}{\gamma(x)} \theta_{101} (x) - \{ \delta_0 (x)\xi_{01} (x)\\
				&~~~~~~+ \{1 - \delta_0(x)\} \xi_{00} (x)\}\bigg]\psi(x).
			\end{align*}
			
			Next, we demonstrate that when $\xi_{10}(x) = 0$, $\theta_{101} (x)= 1$, and $\xi_{00}(x) = 1$, the lower bound is attainable. The key point is to show that there exists a compatible distribution that satisfies all these conditions. Consider the following distributions:
			\begin{align*}
				\xi_{10}(x) &= f(Y = 1 \mid Z = 1, S = 1, R = 0,x),\\
				\theta_{101} (x)&=f\left( Y(1) = 1 \mid S(0) = 0, S(1) = 1,X =x\right)\\
				&=f\left( Y = 1 \mid Z=1, G=\LD,X =x\right),\\
				\xi_{00}(x) &= f(Y = 1 \mid Z = 0, S = 1, R = 0,x).
			\end{align*}
			
			The parameter $\xi_{10}(x)$ and $\theta_{101}(x)$ are conditioned on the treatment group, while $\xi_{00}(x)$ is conditioned on the control group, so there is no conflict between $(\xi_{10}(x),\theta_{101}(x))$ and $\xi_{00}(x)$. Additionally,    to ensure that $\xi_{10}(x) = 0$ and $\theta_{101}(x) = 1$, we can set
			\begin{align*}
				f(Y = 1 \mid Z = 1, G = \LL, R = 0,x)&=0,\\
				f(Y = 1 \mid Z = 1, G = \LD, R = 0,x)&=0,\\
				f(Y = 1 \mid Z = 1, G = \LD, R = 1,x)&=1,\\
				f(R = 1 \mid Z = 1, G = \LD, x)&=1.
			\end{align*}
			There are no  conflicts among these equations.
			When $ \Delta_{\LL}^l(x) $ is attainable for any $ X = x $, it naturally follows that $ \Delta_{\LL}^l $ is attainable, indicating that the lower bound can be achieved.
			The proof for the upper bound follows similarly.

		\end{proof}
		%With the proof of the sharp bound in randomized trials, the proposition follows by applying the previous result conditionally on $X = x$. For simplicity, we omit the detailed derivations.
		
\section{Additional simulation studies}
\label{sec:cov-simu}
In this section,  we   extend our simulation studies to include scenarios with mixed covariate types. Specifically,   we now include a case with mixed continuous and binary covariates: $ A $ is generated as a binary variable from a Bernoulli distribution with probability 0.3 ($ A \sim \text{Bern}(0.3) $), while $ C $ is generated as a continuous variable from a uniform distribution over $[0,1]$ ($ C \sim \text{Unif}[0,1] $). The data generation mechanisms for other variables remain consistent with those described in Section \ref{sec:sim-settings} of the main text. 

The estimation results for this scenario are presented in Table \ref{tab:simu-mixed}. The simulation results presented in Table \ref{tab:simu-mixed} are similar to those in Table \ref{tab:simu} of the main text. Both demonstrate that the proposed estimation method outperforms existing approaches in terms of bias,  RMSE, and 95\% coverage rate across different sample sizes. 

\begin{table}[t]
	\centering
	\caption{Bias ($\times 100$), root mean square error (RMSE) ($\times 100$), and   95\%  coverage probability ($\times 100$) of estimating $\Delta_{\LL}$ under different sample sizes. }
	\label{tab:simu-mixed}
	\resizebox{0.94849\textwidth}{!}{
		\begin{tabular}{ccccc} 
			\toprule                 			Estimation Method                               & Sample Size & Bias ($\times 100$) & RMSE ($\times 100$) & 95\% Coverage Probability ($\times 100$)  \\ 
			\midrule
			\multirow{3}{*}{The naive estimator}                         
			& 500        & $-3.50$               & 7.84                & 92.0 
			\\                 
			& 2000        & $-2.61$               & 4.35                & 88.0                         \\
			& 5000        & $-2.64$               & 3.38              & 79.0                         \\ 
			\midrule
			\multirow{2}{*}{ The method proposed~}                                                        
			& 500        & 73.20              & 95.65              & 35.9                          \\
			\multirow{2}{*}{ in Wang et al. (2017)}& 2000        & 74.98              & 75.35              & 1.1                          \\
			& 5000        & 74.53              & 74.65              & 0                            \\ 
			\midrule
			\multirow{3}{*}{The   proposed estimator} 
			& 500        & -0.40               & 12.30               & 94.8                         \\
			& 2000        & -0.08               & 10.75               & 93.9                         \\
			& 5000        & 0.08          & 6.40               & 93.0\\
			\bottomrule
		\end{tabular}
	}
\end{table}
  \section{Additional real data analyses}
\label{sec:thre}
{In this section, we explore the impact of the threshold used to define the binary outcome variable \( Y \). In the main analysis, we define \( Y = 1 \) if the change in CD4 cell count (i.e., the difference between the CD4 level at 96 weeks and the baseline level) is greater than 0, and \( Y = 0 \) otherwise. To assess the robustness of our findings, we consider alternative threshold values, denoted by \( \kappa \), and redefine \( Y = 1 \) if the CD4 change exceeds \( \kappa \), and \( Y = 0 \) otherwise. The case \( \kappa = 0 \) corresponds to the outcome definition used in the main analysis. 
	
	We apply the estimation procedure introduced in Section~\ref{Sec:Methodology} for analysis. The resulting point estimates and 95\% confidence intervals for varying values of \( \kappa \) are presented in Figure~\ref{fig:etazaresults2}.  
	As shown in this figure, although the point estimates of the causal effect vary under different choices of the threshold \(\kappa\), all of the corresponding 95\% confidence intervals contain zero. This suggests that the estimated treatment effects are not statistically significant under the threshold values considered. In other words, treatment assignment does not significantly affect  the binary outcome, regardless of how the outcome is defined in terms of CD4 count improvement. This finding supports  the robustness of our conclusions to variations in the threshold used to define the outcome.
	
	\begin{figure}[t]
		\centering
		\includegraphics[width=0.87\linewidth]{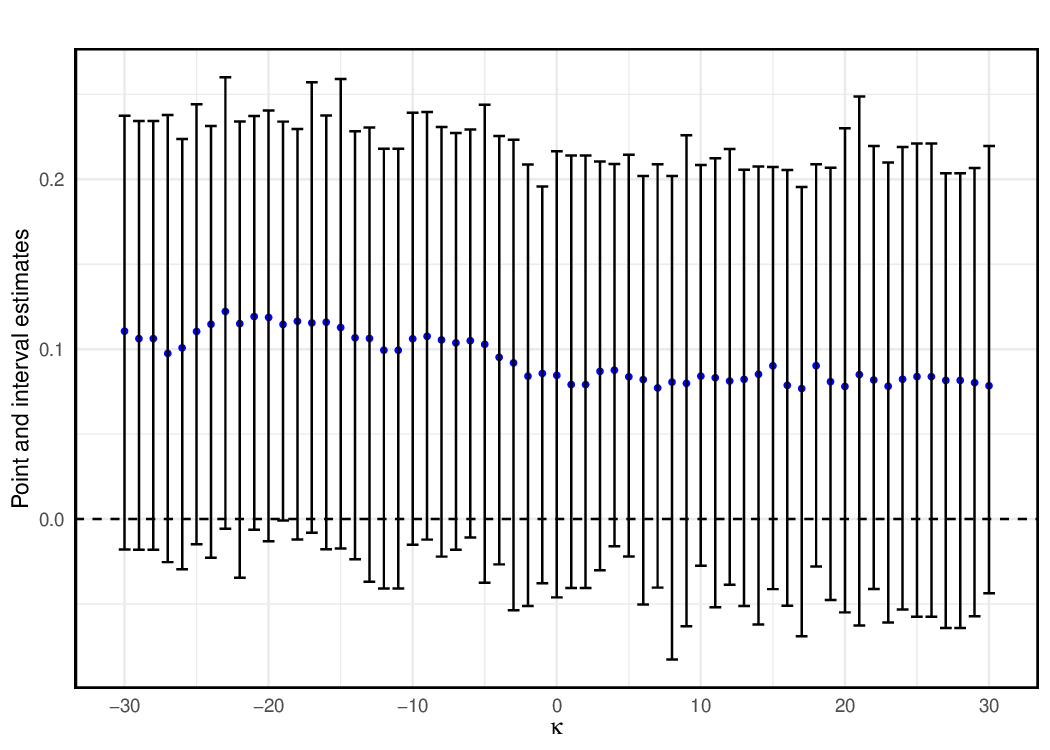}
		\caption{Point estimates under different threshold values for binary outcome definition.}
		\label{fig:etazaresults2}
\end{figure} }

\section{Extension with intermediate variables}
\label{sec:extension}

In this section, we introduce an approach that incorporates intermediate variables to relax Assumption \ref{assu:proxy}(i). Let $ M$ represent an intermediate variable between treatment $ Z$ and outcome $ Y$, which may also influence survival status $ S$. Let $M(z)$ denote the potential value of the intermediate variable for $z=0,1$.
Our goal remains to identify the survivor average causal effect, defined as    $\Delta_{\LL}=\mathbb{E}\{Y(1)-Y(0) \mid G=\LL\}$.
To achieve identification of $ \Delta_{\LL}$, we retain Assumption \ref{assu:mono} from the main text. However, we modify Assumptions \ref{assu:igno}, \ref{assu:shadow}--\ref{assu:proxy} by incorporating the intermediate variable $M$ as follows. 

\begin{assumption}[Strong ignorability and overlap] 
	\label{assu:igno-inter}
	(i) $
	Z \indep \{Y(1),Y(0),S(1),S(0),\\M(1),M(0)\}\mid {X}
	$; (ii) $0<f(Z=1\mid X)<1$.
\end{assumption} 

\setcounter{assumption}{2}
\begin{assumption}
	[Treatment-independent missingness]
	(i) $ Z  \nindep Y \mid (S=1, {X},M)$, (ii) $Z \indep R  \mid (S = 1, {X},M,Y).$  
	\label{assu:shadow-inter}
\end{assumption}

\begin{assumption}[Proxy variable] (i) $
	A \indep Y\mid (Z=1,G,C,M)$, (ii) $A \nindep G\mid (Z=1,S=1,C,M). %%%确认一下
	$
	\label{assu:proxy-inter}
\end{assumption}

Instead of assuming strict orthogonality in Assumption \ref{assu:proxy}, Assumption \ref{assu:proxy-inter} relaxes this condition by introducing intermediate variable $ M$, which captures how $ A$ affects $ Y$ through indirect pathways. However, Assumptions \ref{assu:igno-inter}, \ref{assu:mono} and \ref{assu:shadow-inter}--\ref{assu:proxy-inter} alone are not sufficient to ensure the identification of $ \Delta_{\LL}$. To address this, we introduce the following principal ignorability assumption for the intermediate variable.

\begin{assumption}[Principal ignorability for intermediate variable]\label{assu:prin}
	$f\{M(1)\mid G=\LL,X\}=f\{M(1)\mid G=\LD,X\}.$
\end{assumption}

Assumption \ref{assu:prin} extends the principal ignorability assumption for the outcome, as proposed in \cite{ding2017principal} and \cite{jiang2022multiply}, to the intermediate variable. It states that, given observed covariates, the distribution of the potential intermediate variable remains the same across principal strata of survivors. In other words, $ X$ accounts for all confounding between the potential values of the intermediate variable and survival status.  
Under Assumption \ref{assu:igno-inter} and Assumption \ref{assu:prin},  we obtain
\begin{align} f(M(1) \mid G=\LL, X) = f({M(1) \mid G=\LD, X}) = f(M \mid Z=1, S=1, X). \label{M-equ} \end{align}
The simplest causal graph associated with Assumptions \ref{assu:igno-inter}-\ref{assu:prin} is provided in Figure \ref{fig:causal-inter}.

\begin{figure}[h]
	\centering
	\begin{tikzpicture}
		
		% Define nodes at specific coordinates with increased size
		\node[circle, draw, thick, minimum size=1cm] (A) at (4, 8) {$A$};
		\node[circle, draw, thick, minimum size=1cm] (Z) at (0, 6) {$Z$};
		\node[circle, draw, fill=gray!15, thick, minimum size=1cm] (G) at (6, 6) {$G$};
		\node[circle, draw, thick, minimum size=1cm] (M) at (4, 4) {$M$};
		\node[circle, draw, thick, minimum size=1cm] (S) at (0, 2) {$S$};
		\node[circle, draw, thick, minimum size=1cm] (Y) at (4, 2) {$Y$};
		\node[circle, draw, thick, minimum size=1cm] (R) at (8, 2) {$R$};
		
		% Draw directed edges
		\draw[->, -stealth, thick] (A) -- (Z);
		\draw[->, -stealth, thick] (A) -- (G);
		\draw[->, -stealth, thick] (A) -- (M);
		\draw[->, -stealth, thick] (Z) -- (S);
		\draw[->, -stealth, thick] (Z) -- (M);
		\draw[->, -stealth, thick] (G) -- (M);
		\draw[->, -stealth, thick] (G) -- (Y);
		\draw[->, -stealth, thick] (M) -- (Y);
		\draw[->, -stealth, thick] (M) -- (S);
		\draw[->, -stealth, thick] (S) -- (Y);
		\draw[->, -stealth, thick] (G) -- (R);
		\draw[->, -stealth, thick] (Y) -- (R);
		\draw[->, -stealth, thick, bend left=45] (A) to (R);
		\draw[->, -stealth, thick] (M) to (R);
		
	\end{tikzpicture}
	\caption{The node $A$ represents the proxy variable, the node $Z$ represents the treatment variable, the node $G$ represents principal stratification (latent variable), the node $M$ represent  intermediate variables, the node $S$ represents survival status, the node $R$ represents the missingness indicator, and the node $Y$ represents the outcome. We omit the observed covariates $C$ for simplicity.}
	\label{fig:causal-inter}
\end{figure}

\begin{theorem}
	Under Assumption \ref{assu:igno-inter}, Assumption \ref{assu:mono} in the main text, and Assumptions \ref{assu:shadow-inter}--\ref{assu:prin},  
	$\Delta_{\LL}$ is identifiable.
	\label{thm:iden-inter}
\end{theorem}

The outline of the proof for Theorem \ref{thm:iden-inter} can be summarized as follows:
\begin{itemize}
	\item[]{\it Step 1}: 	Given Assumption \ref{assu:shadow-inter},  
	the conditional distributions $  f(R=1\mid   S =1 ,A,C,Y,M)$ and $f(   Y\mid Z,S=1,A,C,M)$ are identifiable.  The proof of this step follows a similar logic to Proof \ref{proof-thm1-1} in Section \ref{sec:proof-thm1}, we omit it for simplicity.   \item[]{\it Step 2}: 	Given Assumption \ref{assu:igno-inter}, Assumption \ref{assu:mono} in the main text, and Assumptions \ref{assu:shadow-inter}--\ref{assu:prin},   
	the conditional distributions: (1) $
	f(R=0, Y \mid Z , S=1, A, C,M) $, (2) $ 
	f(Y \mid Z=0, G=\LL, A, C,M) 
	$, (3) $f(G=g\mid Z=S=1,A,C,M)$ and (4) $ \E(Y \mid Z=1, G=\LL, C,M) $  are identifiable. We provide the identifiability   in Proof \ref{proof-ps-M}.
	\item[]{\it Step 3}: 	Given Assumption \ref{assu:igno-inter}, Assumption \ref{assu:mono} in the main text, and Assumptions \ref{assu:shadow-inter}--\ref{assu:prin}, the conditional expectations (1) $ \E(Y(0) \mid G=\LL)$ and (2) $ \E(Y(1) \mid G=\LL)$   are identifiable. We provide the identifiability in Proof \ref{proof-prop-ps}.
\end{itemize}
\begin{proof}
	\label{proof-ps-M}
	We proceed with the proof as follows:
	
	\begin{itemize}
		\item[] \textbf{Proof of (1) in {\it Step 2}:}  By the {\it Step 1}, we know that $ f(Y \mid Z, S=1, A, C,M)$ is identifiable. Additionally, $ f(R=0 \mid Z, S=1, C, Y,M) = 1 - f(R=1 \mid Z, S=1, C, Y,M)$ is also identifiable. Therefore:
		$$
		f(R=0, Y \mid Z, S=1, A, C,M) = f(Y \mid Z, S=1, A, C,M) f(R=0 \mid Z, S=1, C, Y,M)
		$$
		is identifiable.
		
		%  \item[] \textbf{Proof of (2):}  
		% From the first step, $ f(Y \mid Z, S=1, A, C)$ is identifiable.
		
		\item[] \textbf{Proof of (2) in {\it Step 2}:}  
		Under Assumption \ref{assu:mono}, we have:
		$$
		f(Y \mid Z=0, S=1, A, C,M) = f(Y \mid Z=0, G=\LL, A, C,M),
		$$
		which  implies that    $ 
		f(Y \mid Z=0, G=\LL, A, C,M) 
		$  is identifiable.
		
		\item[] \textbf{Proof of (3) in {\it Step 2}:}   Given Assumptions \ref{assu:igno-inter}, \ref{assu:mono}, \ref{assu:prin} and equation \eqref{M-equ}, we obtain:
		\begin{align}
			\begin{aligned}
				f(G=\LL \mid A, C, M(1)) &= \dfrac{f(M(1) \mid A, C ,G=\LL ) f(G=\LL\mid A,C)}{f(M(1) \mid A, C   )}\\
				&= \dfrac{f(M  \mid Z=1,S=1,A, C  ) f(G=\LL\mid A,C)}{f(M  \mid Z=1,A, C   )}\\
				&= \dfrac{f(M  \mid Z=1,S=1,A, C  ) f(S=1\mid Z=0,A,C)}{f(M  \mid Z=1,A, C   )}.
			\end{aligned} 
			\label{eq:GLL-M}
		\end{align}
		The proportion $ f(G=\LL \mid Z=S=1, A, C,M)$ can be identified using:
		\begin{align*}
			f&(G=\LL \mid Z=S=1, A, C,M) \\& = \frac{f(G=\LL \mid Z=1, A, C,M)}{f(S=1 \mid Z=1, A, C,M)}  \\& = \frac{f(G=\LL \mid Z=1, A, C,M(1))}{f(S =1 \mid Z=1, A, C,M )}   \tag{Consistency}   \\& = \frac{f(G=\LL \mid  A, C,M(1))}{f(S =1 \mid Z=1, A, C,M )}   \tag{Assumption \ref{assu:igno-inter}} \\
			&=  \dfrac{f(M  \mid Z=1,S=1,A, C  ) f(S=1\mid Z=0,A,C)}{f(M  \mid Z=1,A, C   ){f(S=1 \mid Z=1, A, C,M)}}  \tag{Due to  \eqref{eq:GLL-M}}  \\
			&=\dfrac{f(S=1\mid Z=0,A,C)}{f(S=1 \mid Z=1, A, C)}    
		\end{align*}
		Consequently, $ f(G=\LD \mid Z=S=1, A, C,M)$ is identified as:
		$$
		f(G=\LD \mid Z=S=1, A, C,M) = 1 - f(G=\LL \mid Z=S=1, A, C,M).
		$$
		\item[] \textbf{Proof of (4) in {\it Step 2}:}  
		Under Assumption \ref{assu:proxy-inter}(i), by direct calculation, we have:  
		\begin{align*}
			f&(Y \mid Z=1, S=1, A=a_1, C,M)\\ &= f(Y \mid G=\LL, Z=1, C,M) f(G=\LL \mid Z=1, S=1, A=a_1, C,M) \\
			&\quad + f(Y \mid G=\LD, Z=1, C,M) f(G=\LD \mid Z=1, S=1, A=a_1, C,M), \\
			f&(Y \mid Z=1, S=1, A=a_2, C,M) \\&= f(Y \mid G=\LL, Z=1, C,M) f(G=\LL \mid Z=1, S=1, A=a_2, C,M) \\
			&\quad + f(Y \mid G=\LD, Z=1, C,M) f(G=\LD \mid Z=1, S=1, A=a_2, C,M),
		\end{align*}
		where $ a_1 \neq a_2$.  
		
		Since the coefficient matrix:
		$$
		\begin{bmatrix}
			f(G=\LL \mid Z=1, S=1, A=a_1, C,M) & f(G=\LD \mid Z=1, S=1, A=a_1, C,M) \\
			f(G=\LL \mid Z=1, S=1, A=a_2, C,M) & f(G=\LD \mid Z=1, S=1, A=a_2, C,M)
		\end{bmatrix}
		$$
		is of full rank under Assumption \ref{assu:proxy-inter}(ii), the identifiability of $f(Y \mid Z=1, G=\LL, C,M)=f(Y \mid Z=1, G=\LL,A, C,M)$  can be achieved.
	\end{itemize}
\end{proof} 
\begin{proof}
	\label{proof-prop-ps}
	We proceed with the proof as follows:
	
	\begin{itemize}
		\item[] {\bf Proof of (1) in {\it Step 3}:}   By the {\it Step 2}, we know that $ f(Y \mid Z=0, G=\LL, A, C, M)$ is identifiable. Under Assumption \ref{assu:igno-inter}, we have
		\begin{align*}
			f(Y \mid Z=0, G=\LL, A, C, M) 
			&= f(Y(0) \mid Z=0, G=\LL, A, C, M(0))  \\
			&= f(Y(0) \mid G=\LL, A, C, M(0)).
		\end{align*}
		Thus, $ f(Y(0) \mid G=\LL, A, C, M(0))$ is identifiable.
		
		Next, we show the identifiability of $ f(Y(0) \mid G=\LL)$:
		\begin{align*}
			f&(Y(0) \mid G=\LL) \\
			&= \int f(Y(0), A, C, M(0) \mid G=\LL) \, d\mu(A) \, d\mu(C) \, d\mu(M(0)) \\
			&= \int f(Y(0) \mid A, C, M(0), G=\LL) f(A, C, M(0) \mid G=\LL) \, d\mu(A) \, d\mu(C)\, d\mu(M(0)).
		\end{align*}
		
		Noting that:
		\begin{align*}
			f(A, C, M(0) \mid G=\LL) &= \frac{f(G=\LL \mid A, C, M(0)) f(A, C, M(0))}{\int f(G=\LL \mid A, C, M(0)) f(A, C, M(0)) \, d\mu(A) \, d\mu(C)\, d\mu(M(0))}\\&\propto {f(G=\LL \mid A, C, M(0)) f(A, C, M(0))},
		\end{align*}
		it suffices to identify $ f(G=\LL \mid A, C, M(0))$ and $ f(A, C, M(0))$. Given Assumption \ref{assu:igno-inter}, $ f(A, C, M(0))$ is identifiable. Given Assumptions \ref{assu:mono} and \ref{assu:prin}, we have
		\begin{align*}
			f(G=\LL \mid A, C, M(0)) &= f(G=\LL \mid Z=0, A, C, M )\\
			&= f(S=1 \mid Z=0, A, C, M).
		\end{align*}The identifiability of \( f(Y(0) \mid G=\LL) \) naturally ensures the identifiability of \( \E(Y(0) \mid G=\LL) \).
		Therefore, $ \E(Y(0) \mid G=\LL)$ is identifiable.
		
		\item[] {\bf Proof of (2) in {\it Step 3}:}   We begin by noting:
		\begin{align*}
			f(Y \mid Z=1, G=\LL,A, C, M) 
			&= f(Y(1) \mid Z=1, G=\LL, A, C, M(1))  \\
			% &= \frac{f(Y(1), G=\LL \mid Z=1, A, C, M(1))}{f(G=\LL \mid Z=1, A, C, M(1))} \\
			% &= \frac{f(Y(1), G=\LL \mid A, C, M(1))}{f(G=\LL \mid A, C, M(1))} \\
			&= f(Y(1) \mid G=\LL, A, C, M(1)).
		\end{align*}
		Thus, $ f\{Y(1) \mid G=\LL, A, C, M(1)\}$ is identifiable.
		
		Next, we show the identifiability of $ f(Y(1) \mid G=\LL)$:
		\begin{align*}
			f&(Y(1) \mid G=\LL)\\ 
			&= \int f(Y(1), A, C, M(1) \mid G=\LL) \, d\mu(A) \, d\mu(C)\, d\mu(M(1)) \\
			&= \int f(Y(1) \mid A, C, M(1), G=\LL) f(A, C, M(1) \mid G=\LL) \, d\mu(A) \, d\mu(C)\, d\mu(M(1)).
		\end{align*}
		
		Noting that:
		$$
		f(A, C, M(1) \mid G=\LL) \propto{f(G=\LL \mid A, C, M(1)) f(A, C, M(1))} ,
		$$
		it suffices to identify $ f(G=\LL \mid A, C, M(1))$ and $ f(A, C, M(1))$. Given Assumption \ref{assu:igno-inter}, $ f(A, C, M(1))$ is identifiable. We have   established the identifiability of \( f(  G=\LL \mid A, C, M(1)) \) in  \eqref{eq:GLL-M}. These results  guarantee  the identifiability of \( \E(Y(1) \mid G=\LL) \). 
	\end{itemize}  
\end{proof}
\end{document}